\documentclass[aps,prc,article,nofootinbib,superscriptaddress]{revtex4-2} 
\usepackage[utf8]{inputenc}
\usepackage{graphicx,comment}
\usepackage{float}
\usepackage{units}
\usepackage{multirow,array}
\usepackage{xcolor}
\usepackage{amsmath,amssymb}
\usepackage{yfonts}
\usepackage{tensor}
\usepackage{paralist}
\usepackage{braket}
\usepackage{tensor,mathrsfs}
\usepackage[normalem]{ulem}
\usepackage{caption}
\usepackage{subcaption}
\usepackage[hidelinks]{hyperref}

\newcommand{\dd}{\mathrm{d}}

\newcommand{\avept}{$\langle p_\mathrm{T} \rangle$}
\newcommand{\nch}{$N_\mathrm{ch}$}
\newcommand{\etsum}{$E_\mathrm{T,sum}^\mathrm{HF}$}
\newcommand{\pt}{$p_\mathrm{T}$}

\newcommand{\uaAff}{\affiliation{Department of Physics and Astronomy, University of Alabama, 514 University Boulevard, Tuscaloosa, AL 35487, USA}}
\newcommand{\uuAff}{\affiliation{Institute for Theoretical Physics, Utrecht University, Princetonplein 5, Utrecht, 3584 CC, Netherlands}}

\graphicspath{figs_Update}

\begin{document}
\title{Anisotropic time evolution of sound modes in Bjorken expanding holographic plasma}

\author{Casey Cartwright}
\email{c.c.cartwright@uu.nl}
\uuAff

\author{Ruchi Chudasama}
\email{rchudasama@ua.edu}
\uaAff

\author{Sergei Gleyzer}
\email{sgleyzer@ua.edu}
\uaAff

\author{Durdana Ilyas}
\email{dilyas@crimson.ua.edu}
\uaAff

\author{Matthias Kaminski}
\email{mski@ua.edu}
\uaAff

\author{Marco Knipfer}
\email{mknipfer@crimson.ua.edu}
\uaAff

\author{Jun Zhang}
\email{jzhang163@crimson.ua.edu}
\uaAff
\date{\today}

\begin{abstract}
The speed of sound is a key parameter for characterizing equilibrium states. However, sound waves change their properties when propagating through rapidly evolving anisotropic media, such as the quark-gluon plasma created in heavy-ion collisions. 
This paper uses $\mathcal{N}=4$ Super-Yang-Mills theory to numerically study the time evolution of the speed and attenuation of sound modes along with the relaxation time in a plasma undergoing Bjorken expansion from various initial states in a quasi-static approximation. 
The longitudinal Bjorken expansion breaks the isotropy, resulting in two distinct sound speeds that range from just below the conformal value to the speed of light. 
An anisotropic hydrodynamic description is constructed and its applicability is discussed.   
Implications for the analysis of heavy ion data are considered. 
\end{abstract}
\maketitle
\tableofcontents
%

\section{Introduction} 
\label{sec:intro} 
%
%
In the analysis of experimental data from heavy-ion collisions conducted at LHC or RHIC it is often assumed that the generated quark-gluon plasma is effectively isotropic and in (local) thermodynamic equilibrium over a large part of its evolution; direction and time dependent transport coefficients are often replaced with isotropic time-averages. However, several effects are known to make the plasma strongly anisotropic.\footnote{The anisotropy stems from the initial state (anisotropic initial state geometry, the early-time momentum-space anisotropy, initial magnetic field, etc.) while the equations of motion are isotropic.} Among those effects there are, for example, the initial spatial anisotropy of the asymmetric overlap region of the colliding ions leading to elliptic flow~\cite{Ollitrault:1992bk,PHENIX:2004vcz,BRAHMS:2004adc,Adams:2005dq,PHOBOS:2004zne,ALICE:2013xna,ATLAS:2015hzw}, and the initially strong magnetic field generated by the colliding ions~\cite{Deng:2012pc,McLerran:2013hla,Roy:2015kma,Stewart:2021mjz,Yan:2021zjc,Adhikari:2024bfa}. Furthermore, the plasma is never in global thermodynamic equilibrium, at best it may reach local thermodynamic equilibrium but no earlier than 1~femtosecond after the collision~\cite{Kolb:2000fha,Huovinen:2001cy,Heinz:2004pj,Chesler:2008hg,Chesler:2010bi,Gale:2013da}. Most prominently, the experimentally well-established\footnote{The Bjorken expanding fluid flow describes a boost-invariant plasma expansion. The fact that the quark gluon plasma generated in heavy-ion collisions~\cite{back2003significance,chatrchyan2011dependence} expands boost-invariantly can be deduced from the central plateau, {i.e.,} the plateau at mid-rapidity values in particle multiplicity distributions as function of the rapidity.} boost-invariant Bjorken-expansion~\cite{Bjorken:1982qr} of the plasma leads to a strong anisotropy between the beamline and the transverse plane, while it renders the plasma out of global thermodynamic equilibrium at all times. 
While the assumptions of isotropy and equilibrium considerably simplify the data analysis, they can lead to drastically wrong conclusions about the (thermodynamic and hydrodynamic) properties of the quark-gluon plasma, as we demonstrate in this work. 
\begin{figure}[hbt!]
    \includegraphics[width=0.49\textwidth]{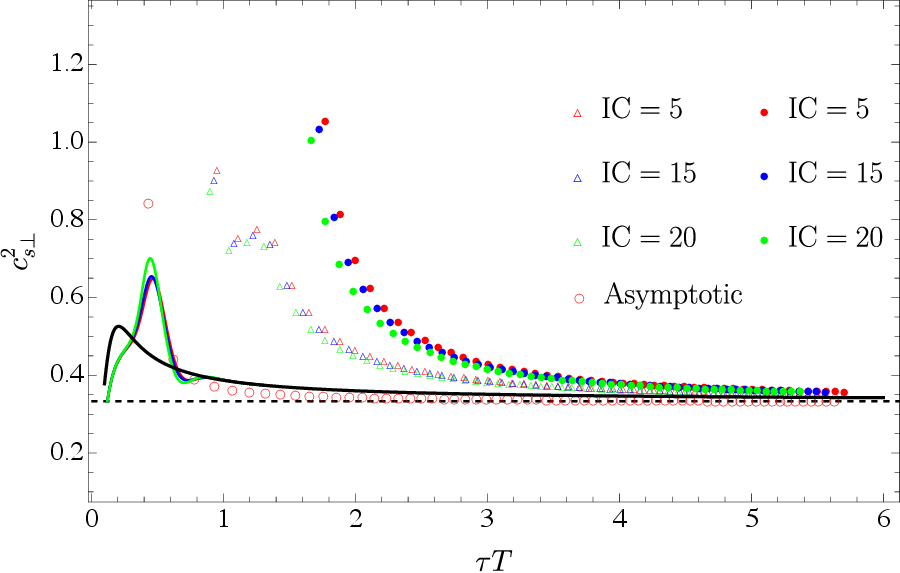}
    \hfill
    \includegraphics[width=0.49\textwidth]{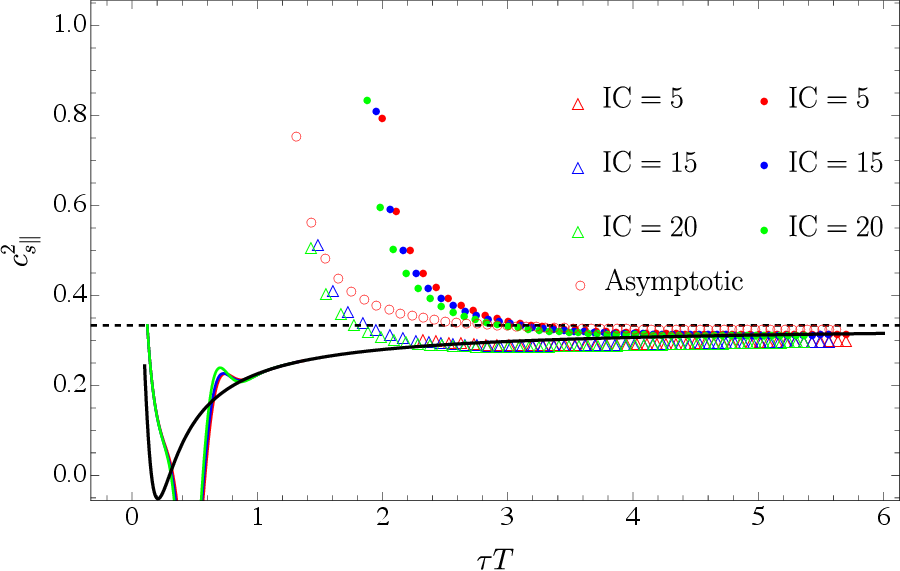}
    \caption{
    {\it Time evolution of the two speeds of sound.} The left (right) figure shows the square of the transverse (longitudinal) speed of sound, $c_{s\perp}^2$ ($c_{s\parallel}^2$). Different colors represent different initial conditions~(IC=5, 15, 20), see table~\ref{tab:IC}. Dashed black lines indicate the conformal value of 1/3, and the solid black curves are the sound attractors computed from a thermodynamic definition of the speed of sound~\cite{Cartwright:2022hlg}. Solid curves in red, blue, and green show the speed of sound computed for the three initial conditions from that same thermodynamic definition of the speed of sound~\cite{Cartwright:2022hlg}. The triangles (dots) correspond to calculations with (without) time-derivative corrections. The red hollow circles display the corresponding results when using the time-dependent asymptotic metric~\eqref{ASP:metric}~\cite{janikAsymptoticPerfectFluid2006,Chesler:2009cy}. 
    The quasi-static approximation and fit procedure break down for the transverse (longitudinal) speed of sound including time-derivatives indicated by hollow triangles around $\tau T\lesssim 1.2$ ($\tau T\lesssim 1.4$). 
    For an estimate of errors, see Figs.~\ref{figSET} and~\ref{figSEL}. 
    }
     \label{figBSS}
\end{figure}

In order to demonstrate how anisotropy and non-equilibrium affect plasmas at strong coupling, we consider a holographic model. We numerically compute the time evolution of the speed, the attenuation, and relaxation time of sound modes for the $\mathcal{N}=4$ Super-Yang-Mills theory prepared in states far from equilibrium and subsequently undergoing a Bjorken expansion. 
At each instant of the Bjorken time evolution, a sound perturbation is introduced and Fourier transformed. 
This assumes a \emph{quasi-static approximation}, i.e., the sound perturbations occur on a time scale much shorter than that of the Bjorken expansion.  
Then, the speed of sound, the sound attenuation coefficient, 
and the relaxation time 
are extracted from the dispersion relation of the eigenfrequencies of that sound perturbation at that instant. 
%

Previous analyses of the hydrodynamic regime of various anisotropic systems have revealed their common properties~\cite{Ammon:2020rvg,Garbiso:2020puw,Endrodi:2018ikq,Ammon:2017ded,Janiszewski:2015ura,Cartwright:2019opv,Cartwright:2021maz,Cartwright:2021qpp,Cartwright:2021hpv,Abbasi:2022aao,Amano:2022mlu,Amano:2023bhg,Kaminski:2023yut,Ghosh:2024fkg,Ghosh:2024owm,Shovkovy:2025yvn,Kaminski:2025ika}. For example,  each isotropic transport coefficient splits into one longitudinal and one transverse to the anisotropy. Due to the symmetry breaking, the corresponding hydrodynamic perturbations will generally satisfy distinct dynamical equations coupling them to distinct symmetry sectors. For example, in holographic models, the longitudinal shear perturbations dynamically couple to the momentum diffusion modes (vector modes) and the longitudinal shear viscosity drops below the conjectured KSS bound~\cite{Policastro:2001yc,kss}, $\eta_\parallel/s<1/(4\pi)$, with the entropy density $s$. Meanwhile, the transverse shear viscosity (remaining associated with a single uncoupled tensor perturbation) saturates the bound, $\eta_\perp/s=1/(4\pi)$, as discussed in detail in Refs.~\cite{Ammon:2020rvg,Garbiso:2020puw,Amano:2023bhg}.\footnote{Violations of the KSS bound generated by such symmetry breaking within Einstein gravity duals first appeared in Refs.~\cite{Rebhan:2009vc,Erdmenger:2008rm,Erdmenger:2010xm}, while violations of the KSS bound outside of Einstein gravity were pointed out already earlier~\cite{Kats:2007mq,Brigante:2007nu,Brigante:2008gz}, as reviewed in detail in~\cite{Cremonini:2011iq}.} In strongly magnetized relativistic plasma, the two charge conductivities (longitudinal and transverse to the magnetic field) have been computed and extrapolated to QCD quark gluon plasma~\cite{Ghosh:2024fkg,Ghosh:2024owm,Shovkovy:2025yvn}. 
A version of viscous anisotropic hydrodynamics~(\emph{viscous aHydro} or \emph{VAH}) has previously been developed and compared to experimental heavy ion collision data, indicating that VAH has a larger range of applicability than isotropic hydrodynamics~\cite{Strickland:2014pga,Strickland:2017kux,Alqahtani:2017jwl,Ryblewski:2010bs,Ryblewski:2010tn,Ryblewski:2011aq,Ryblewski:2012rr,Florkowski:2012lba,Bazow:2013ifa}.  
Sound modes and the time-evolution of the energy-momentum tensor have previously been computed in kinetic theory approaches~\cite{Du:2023bwi,Du:2023ucs,Ochsenfeld:2023wxz,Kushwah:2024ngr,kurkela2019effective,PhysRevD.102.056003}. 
The speed of sound is often discussed as the parameter controlling the \emph{stiffness} of the the equation of state in the context of heavy ion collisions~\cite{Pratt:2015zsa,Alba:2017hhe,Busza:2018rrf,Spieles:2020zaa,Monnai:2021kgu,An:2021wof,Abbasi:2021rlp} or neutron star mergers~\cite{Oechslin:2006uk,Font:2008,Baiotti:2016qnr}.

Summarizing the main results of this present paper, we find that \emph{due to the longitudinal Bjorken expansion, the $\mathcal{N}=4$ Super-Yang-Mills plasma is anisotropic and its anisotropic hydrodynamic transport is highly relevant. Our theoretical results provide a strong motivation for considering anisotropic plasma properties in experiments and in their analysis.} 
As anticipated from the discussion of previous results above, anisotropy gives rise to two distinct speeds of sound, one longitudinal ($c_\parallel$) to the expansion and one transverse to it ($c_\perp$). 
The two speeds of sound far from equilibrium can deviate significantly from their isotropic equilibrium value of $c_s=1/\sqrt{3}$. Instead, their values range from just below the conformal isotropic equilibrium value $c_s=1/\sqrt{3}$ up to the speed of light $c_s\approx1$, as shown in Fig.~\ref{figBSS}. This perturbative calculation of their time evolution (hollow triangles) qualitatively agrees very roughly with a previous calculation using a thermodynamic definition\footnote{Four definitions were compared where partial derivatives $\partial P_{\perp,\parallel}/\partial \epsilon$ are evaluated either at fixed entropy density, or at fixed apparent horizon area, or on a fixed proper time slice, or using a chain rule; the fixed proper time definition turns out to be most reliable~\cite{Cartwright:2022hlg}.}, $c_{\perp,\parallel}=\partial P_{\perp,\parallel}/\partial \epsilon$, of the speeds of sound~\cite{Cartwright:2022hlg}, see solid curves in Fig.~\ref{figBSS}, specifically the \emph{hydrodynamic sound attractors} given by the solid black curves.\footnote{Hydrodynamic attractors have been discussed for Bjorken expanding plasma before~\cite{Ambrus:2023oyk,Mitra:2020mei,Giacalone:2019ldn,Heller:2013fn,Heller:2014wfa,Heller:2015dha,Spalinski:2017mel}.}
However, the deviation of the perturbatively computed values from the thermodynamically computed values can be as large as a factor of two. 
The agreement between thermodynamically and perturbatively computed longitudinal (transverse) sound speed is good for times later than 2.5 fm/c (3.5 fm/c).\footnote{We adapt the unit matching between gravitational and field theoretic quantities suggested by~\cite{Chesler:2009cy}, where the dimensionless $\tau T=1$ corresponds to $1$ femtosecond divided by the speed of light, 1 fm/c.}  
Moreover, the thermodynamic definition (solid curves) systematically \emph{underestimates} the magnitude of both anisotropic sound speeds out of equilibrium, only overestimating the longitudinal sound speed marginally at late times $\tau T\gtrsim 2.4$.\footnote{Here, we have assumed that the perturbative calculation provides a more accurate result for the speed of sound waves in the medium out of equilibrium than the thermodynamically defined speed of sound. } 
Meanwhile, despite the large anisotropy between longitudinal and transverse direction in the Bjorken-expanding plasma, both of the sound attenuation coefficients barely deviate from their isotropic equilibrium value, at most by approximately $10$\% as seen in Fig.~\ref{figSAT}. The two relaxation times evolve very similar to each other and deviate significantly from their isotropic equilibrium values. Around $\tau T\approx 2.5$, the relaxation becomes instantaneous, that is, if the high order fit requiring the third order in the derivative expansion of the dispersion relation can be trusted, see~Fig.~\ref{figRT} and discussion in Sec.~\ref{sec:Holographic description}. 

We begin in the next section, Sec.~\ref{sec:summaryCMSAnalysis}, with a review of the recent CMS analysis extracting a speed of sound value from collision data at the LHC~\cite{CMS:2024sgx}. 
Our construction of the anisotropic hydrodynamic description of sound modes is discussed in Sec.~\ref{sec:hydro}, while our holographic calculation is presented in Sec.~\ref{sec:Holographic description}.\footnote{It should be stressed that our approach to constructing \emph{anisotropic hydrodynamics} systematically differs from that previously developed and referred to as \emph{aHydro}~\cite{Strickland:2014pga,Strickland:2017kux,Ryblewski:2010bs,Ryblewski:2010tn,Ryblewski:2011aq,Ryblewski:2012rr,Florkowski:2012lba,Bazow:2013ifa}.} 
More details regarding these calculations are collected in the appendices~\ref{sec:BjorkenMetric}, \ref{appedix:holographic results for equilibrium case}, \ref{INCs}, \ref{Holographic fluctuation equations}, \ref{sec:StandardErrorFit}. 
For comparison and to test the calculation methods, the speed, attenuation and relaxation time of sound in the analytically known isotropically expanding Vaidya plasma were computed, see appendix~\ref{sec:SoundVaidya}. 
We discuss the implications of our results for experimental measurements of sound properties such as~\cite{CMS:2024sgx} in the Conclusions, Sec.~\ref{sec:conclusions}.
%

\section{Review of the CMS analysis extracting a speed of sound} 
\label{sec:summaryCMSAnalysis} 
In order to illustrate the impact of our results on experiments, we now review a recent experimental analysis yielding a speed of sound.  
%
The CMS Collaboration has recently reported a determination of the speed of sound in the quark-gluon plasma (QGP) created in ultra-central PbPb collisions at 
$\sqrt{s_{\rm {NN}}}=5.02$ TeV~\cite{CMS:2024sgx}, using data collected in 2018, with an integrated luminosity of 0.607 nb$^{-1}$. 

The method leveraged a hydrodynamic probe focusing on ultra-central collisions (near-zero impact parameter $b \approx 0$), where the nuclear radius fixed the system size, and increases in charged-particle multiplicity (\nch) arose from fluctuations in parton interactions, leading to higher entropy density ($s$) at nearly constant initial energy density and temperature. This was used to calculate the speed of sound squared ($c^2_{\mathrm{s}}$) from thermodynamic relations as
\begin{equation}
\label{eq:sound2}
c^2_{\mathrm{s}}
= \frac{\mathrm{d}P}{\mathrm{d}\varepsilon}
= \frac{s\,\mathrm{d}T}{T\,\mathrm{d}s}
= \frac{\mathrm{d}\langle p_\mathrm{T} \rangle / \langle p_\mathrm{T}\rangle}
       {\mathrm{d}\langle N_\mathrm{ch}\rangle / \langle N_\mathrm{ch}\rangle}. 
\end{equation}
 Thus, the analysis measures the experimental observable mean transverse momentum (\avept) of charged particles in the event as a function of charged-particle multiplicity\nch, where both quantities are measured within the same kinematic ranges. The centrality of PbPb collisions, which quantifies the degree of nuclear overlap or the impact parameter, was determined using the total transverse energy deposited in both hadron forward (HF) calorimeters denoted as (\etsum), which is linearly related to \nch. The analysis was restricted to the 10\% most central events, corresponding to the highest \etsum values (3400-5200 GeV), thereby selecting ultra-central collisions characterized by an impact parameter approaching zero. Charged particle tracks reconstructed with \pt$>0.3$ GeV within $|\eta|<0.5$ were utilized and corrected for efficiency and misreconstruction using HYDJET simulations. Further, \pt~spectra were extrapolated to the full range of \pt~down to 0 GeV via the Hagedorn function fits over 0.4-4.5 GeV. Both \avept and \nch were normalized to the 0-5\% centrality class with \avept$_0 \approx 658 \pm 25$ MeV, corresponding to $T_{\mathrm{eff}} = 219 \pm 8$ MeV). 
 
 The multiplicity distribution was fitted to constrain the “knee” parameters (\nch$^\mathrm{knee} \approx 1.11$, $\sigma \approx 0.0272$), and a hydrodynamically motivated power-law form was applied to the high-multiplicity tail (\nch$^{\mathrm{norm}}>1.14$), yielding $c_s^2 = 0.241 \pm 0.002$ (stat) $\pm 0.016$ (syst). Systematic uncertainties are reported to be dominated by tracking corrections, $p_\mathrm{T}$ extrapolation, and fit-range variations. The observed rise of \avept~with multiplicity is consistent with hydrodynamic models such as TRAJECTUM ($c_s^2 \approx 0.283 \pm 0.045$) and the framework of Gardim et al.~\cite{Gardim:2019brr}, and agrees precisely with lattice QCD predictions at $T\approx219$ MeV. 
This analysis was stated to provide stringent constraints on the QGP equation of state and strong evidence for the existence of a deconfined QCD phase at LHC energies. 

Subsequently, this CMS analysis~\cite{CMS:2024sgx} was discussed by Gavassino et al.~\cite{Gavassino:2025bts}. In~\cite{Gavassino:2025bts}, it is stated that the argument based on energy and entropy made in~\cite{CMS:2024sgx} yields the ratio of pressure and energy density rather than the speed of sound squared. Furthermore, based on numerical simulations, \cite{Gavassino:2025bts} assert that $\langle p_\mathrm{T}\rangle$ and $N_{\mathrm{ch}}$ are not sufficiently accurate proxies for the energy and entropy. Gavassino et al.~\cite{Gavassino:2025bts} state that the analysis of~\cite{CMS:2024sgx} is strongly dependent on whether the effective volume is strictly constant or not, a condition that could not be enforced experimentally. 

A major problem for any experimental analysis is that the \emph{quark gluon plasma is not an equilibrium system}, for the earliest times there is not even local equilibrium established. 
Results from~\cite{Cartwright:2022hlg} demonstrate this same fact explicitly for the holographic system (considered in this present work in Sec.~\ref{sec:Holographic description}). 
All measures of local thermodynamic equilibrium defined in~\cite{Cartwright:2022hlg} (for example the local validity of the thermodynamic relation $\epsilon+P=s T$ between energy density $\epsilon$, pressure $P$, entropy $s$, and temperature $T$) demonstrate that local equilibrium is not reached before the proper time $\tau\approx 1/T$. For later times $\tau > 1/T$, all measures indicate that local thermodynamic equilibrium is reached. 
As our results in this present work will suggest, \emph{even if there is local equilibrium established, that does not necessarily imply the equilibrium definition of the speed of sound, $c_{s}^2=\partial P/\partial \varepsilon$ to be valid}, see Sec.~\ref{sec:Holographic description} and the Conclusions~\ref{sec:conclusions}. 
Another major challenge for any experimental analysis is that it should take into account the anisotropy of the system. Our results will imply that \emph{anisotropies are highly relevant for the interpretation of hydrodynamic and thermodynamic quantities}, see Sec.~\ref{sec:Holographic description} and~\ref{sec:conclusions}.  
Further implications of our results for experiments and their analysis are mentioned in the Conclusions, Sec.~\ref{sec:conclusions}.

\section{Hydrodynamic perturbations of anisotropic fluids}
\label{sec:hydro} 
%
In this section, we construct the anisotropic hydrodynamic description which we will later use to analyze the numerical data from our anisotropic holographic plasma. Our constitutive relation mirrors the one constructed in~\cite{Ammon:2020rvg}, where the anisotropy was generated by an external magnetic field.\footnote{Our construction here is systematically different from that used in Refs.~\cite{Strickland:2014pga,Strickland:2017kux,Ryblewski:2010bs,Ryblewski:2010tn,Ryblewski:2011aq,Ryblewski:2012rr,Florkowski:2012lba,Bazow:2013ifa}, where the authors consider an anisotropic one-particle distribution function in the context of kinetic theory.} Throughout this paper we consider three spatial and one time dimension. 
\subsection{Isotropic fluid}
\label{sec:isotropicFluid}
In the isotropic case, the constitutive equation for the energy-momentum tensor is given by
\begin{equation}
    T^{\mu\nu} = \epsilon\, u^\mu u^\nu + P\, \Delta^{\mu\nu} + \pi^{\mu\nu},
\end{equation}
where the shear stress tensor takes the form
\begin{equation}\label{IsoShearTensor}
    \pi^{\mu\nu} = -\eta\, \sigma^{\mu\nu},
\end{equation}
with
\begin{equation}
\sigma^{\mu\nu} \equiv 2 \left( \nabla^{\langle \mu} u^{\nu \rangle} \right)
 \equiv 2\left( \Delta^{\mu\alpha} \Delta^{\nu\beta} \nabla_{(\alpha} u_{\beta)} - \frac{1}{3} \Delta^{\mu\nu} \nabla_\lambda u^\lambda \right).
\end{equation}
Here, $\epsilon$, $P$, and $\eta$ denote the energy density, pressure, and shear viscosity, respectively.$\nabla_{(\mu}u_{\nu)}\equiv(\nabla_{\mu}u_{\nu}+\nabla_{\nu}u_{\mu})/2.$ The fluid four-velocity $u^\mu$ satisfies the normalization condition $u^\mu u_\mu = -1$. The projector onto the space orthogonal to $u^\mu$ is defined as
\begin{equation}
    \Delta^{\mu\nu} \equiv g^{\mu\nu} + u^\mu u^\nu. 
\end{equation}

\subsection{Anisotropic conformal fluid}
\label{sec:anisotropicFluid}
We now consider a conformal relativistic fluid with anisotropy along a fixed spacelike direction \( n^\mu \), which is orthogonal to the fluid velocity \( u^\mu \). The energy-momentum tensor is expressed as~\cite{Ammon:2020rvg}
\begin{equation}\label{Tconstitutive2}
    T^{\mu\nu} =
\epsilon\, u^\mu u^\nu
+ P_T\, \Xi^{\mu\nu}
+ P_L\, n^\mu n^\nu
+ \pi^{\mu\nu}_\perp
+ \pi^{\mu\nu}_\parallel+n^{\langle \mu} n^{\nu \rangle} \left( \eta_1 \theta_\perp + \eta_2 \theta_\parallel  \right),
\end{equation}
where $P_T$ and $P_L$ represent the transverse and longitudinal pressures, respectively, 
and $\eta_1, \, \eta_2$ are anisotropic shear viscosities, associated with the two shears in the two planes which include the anisotropy direction, namely the $(x_1,x_3)$-plane and the $(x_2,x_3)$-plane~\cite{Ammon:2020rvg}. 
The anisotropy direction $n^\mu$ satisfies \( n^\mu n_\mu = +1 \) and \( u^\mu n_\mu = 0 \). We define the transverse projector orthogonal to both $u^\mu$ and $n^\mu$ as
\begin{equation}
    \Xi^{\mu\nu} \equiv \Delta^{\mu\nu} - n^\mu n^\nu.
\end{equation}

The transverse and longitudinal components of the shear stress tensor are defined as
\begin{equation}\label{TShearTensor}
\pi_{\perp}^{\mu\nu} \equiv -2\eta_{\perp} \left[ \left\langle \nabla^{\mu} u^{\nu} \right\rangle_{\perp} - \frac{1}{2} \Xi^{\mu\nu} \theta_{\perp} \right]=-\frac{1}{2}\eta_{\perp}(\Xi^{\mu\alpha}\Xi^{\nu\beta}+\Xi^{\mu\beta}\Xi^{\nu\alpha}-\Xi^{\mu\nu}\Xi^{\alpha\beta})\sigma_{\alpha\beta},
\end{equation}
and
\begin{equation}\label{LShearTensor}
\pi_{\parallel}^{\mu\nu} \equiv -2\eta_{\parallel} \left[ \left\langle \nabla^{\mu} u^{\nu} \right\rangle_{\parallel} - 2 n^{\mu} n^{\nu} \theta_{\parallel} \right]=-\eta_{\parallel}(n^\mu\Sigma^\nu+n^\nu\Sigma^\mu),    
\end{equation}
respectively. 
Here $\Sigma^\mu\equiv(\Delta^{\mu\nu}-n^\mu n^\nu)\sigma_{\nu\rho}n^\rho$. 
The expansion scalar $\theta$ and its components $\theta_\perp$ and $\theta_\parallel$ are given by :
\[
\theta \equiv \nabla_\lambda u^\lambda, \quad
\theta_\perp \equiv \Xi^{\mu\nu} \nabla_\mu u_\nu, \quad
\theta_\parallel \equiv n^\mu n^\nu \nabla_\mu u_\nu.
\]
These quantities satisfy the following properties:
\begin{align}
    g_{\mu\nu}\!\left\langle\nabla^\mu u^\nu\right\rangle_\perp 
        &= \theta_\perp, \\[4pt]
    g_{\mu\nu}\!\left\langle \nabla^{\mu} u^{\nu} \right\rangle_{\parallel} 
        &= 2\theta_\parallel, \\[4pt]
    \theta_\perp + \theta_\parallel 
        &= \theta. \label{TExpansion}
\end{align}
Equation~\eqref{TExpansion} follows from the identity \( u^\mu u^\nu \nabla_\mu u_\nu = 0 \).
The projected symmetric velocity gradients are
\begin{align}
    \left\langle \nabla^{\mu} u^{\nu} \right\rangle_{\perp}
        &\equiv \frac{1}{2}\,\Xi^{\mu\alpha}\,\Xi^{\nu\beta}
        \left( \nabla_\alpha u_\beta + \nabla_\beta u_\alpha \right),
        \\[6pt]
    \left\langle \nabla^{\mu} u^{\nu} \right\rangle_{\parallel}
        &\equiv \frac{1}{2}
        \left( n^\mu n^\alpha \Delta^{\nu\beta}
             + n^\nu n^\alpha \Delta^{\mu\beta} \right)
        \left( \nabla_\alpha u_\beta + \nabla_\beta u_\alpha \right).
\end{align}

The conformal condition imposes the tracelessness of the energy-momentum tensor, {i.e.},
\begin{equation}
    \epsilon = 2P_T + P_L,
\end{equation} 
and both transverse and longitudinal shear stress tensors are individually traceless, ${\pi^{\mu}_{\mu}}_\perp =0,\, {\pi^{\mu}_{\mu}}_\parallel = 0$, ensuring the tracelessnes of the full energy momentum tensor. 

\subsection{Recovery of the isotropic fluid in the limit of vanishing anisotropy}
In this subsection we demonstrate that the anisotropic hydrodynamic description constructed in Sec.~\ref{sec:anisotropicFluid} reduces correctly to the known isotropic description from Sec.~\ref{sec:isotropicFluid}. In the isotropic limit, we have
\begin{equation}
   \eta_{\perp} = \eta_{\parallel} = \eta, 
\end{equation}
and the expansion rates along each spatial direction satisfy
\begin{equation}
    n_1^\mu n_1^\nu \nabla_\mu u_\nu = n_2^\mu n_2^\nu \nabla_\mu u_\nu = n_3^\mu n_3^\nu \nabla_\mu u_\nu = \theta_\parallel,
\end{equation}
where \( n_1^\mu, n_2^\mu, n_3^\mu \) are three orthonormal spacelike vectors along those three directions. Additionally,
\begin{equation}
    \left( g^{\mu\nu} + u^\mu u^\nu - n_1^\mu n_1^\nu - n_2^\mu n_2^\nu - n_3^\mu n_3^\nu \right) \nabla_\mu u_\nu = 0.
\end{equation}
From these relations, we find
\begin{equation}\label{Isoropic expansion rate}
   2\theta_\parallel = \theta_\perp, \qquad 3\theta_\parallel = \theta.
\end{equation}
Moreover, since
\begin{equation}
    \left\langle \nabla^{\mu} u^{\nu} \right\rangle_{\perp} + \left\langle \nabla^{\mu} u^{\nu} \right\rangle_{\parallel} = \Delta^{\mu\alpha} \Delta^{\nu\beta} \nabla_{(\alpha} u_{\beta)} + n^{\mu} n^{\nu} \theta_\parallel,
\end{equation}
the total shear stress tensor reduces to
\begin{equation}
    \pi^{\mu\nu} = \pi_{\perp}^{\mu\nu} + \pi_{\parallel}^{\mu\nu} = -2\eta \left( \Delta^{\mu\alpha} \Delta^{\nu\beta} \nabla_{(\alpha} u_{\beta)} - \frac{1}{3} \Delta^{\mu\nu} \nabla_\lambda u^\lambda \right),
\end{equation}
which exactly reproduces the shear stress tensor given in Eq.~\eqref{IsoShearTensor} for the isotropic case.
In addition, to match the energy-momentum tensor of isotropic equilibrium systems, the last two terms in Eq.~(\ref{Tconstitutive2}) must vanish, which implies
\begin{equation}\label{eta12}
    2\eta_1+\eta_2=0.
\end{equation}
Equation~(\ref{eta12}) precisely corresponds to the Onsager relations of transport coefficients for equilibrium systems, as discussed in \cite{ammon2021chiral}.\footnote{In \cite{ammon2021chiral}, the corresponding relation is instead $3\eta_1+\eta_2=0$, since there the constitutive relation is defined as 
$T^{\mu\nu} =
\epsilon\, u^\mu u^\nu
+ P_T\, \Xi^{\mu\nu}
+ P_L\, n^\mu n^\nu
+ \pi^{\mu\nu}_\perp
+ \pi^{\mu\nu}_\parallel
+ n^{\langle \mu} n^{\nu \rangle} \left( \eta_1 \theta + \eta_2 \theta_\parallel \right)$.
}

\subsection{Sound dispersion relations in Bjorken expanding plasma}
\label{sec:hydroDispersion}
In this section, we apply the anisotropic hydrodynamic description from Sec.~\ref{sec:anisotropicFluid} to derive the dispersion relations for a specific anisotropic plasma, namely a Bjorken-expanding plasma. One could describe this plasma with a time-dependent fluid velocity defined in Minkowski space. However, it is advantageous to instead choose an expanding coordinate system in which the fluid velocity is trivial. 
With this choice the time-dependent Milne metric in that coordinate system is given by (note the factor $\tau^2$ in the rapidity direction $d\xi^2$) 
\begin{equation}\label{eq:MilneMetric}
    \mathrm{d}s^2 
    = g^{\mathrm{(Milne)}}_{\mu\nu}(\tau)\,\mathrm{d}x^\mu \mathrm{d}x^\nu
    = -\mathrm{d}\tau^2 + \mathrm{d}x_1^2 + \mathrm{d}x_2^2 + \tau^2 \mathrm{d}\xi^2\, ,
\end{equation}
with the proper time $\tau$, and $\xi=\tfrac{1}{2}\ln[(t+x_3)/(t-x_3)]$ is the rapidity in the longitudinal direction (along the Bjorken expansion). 
For the derivation of the dispersion relations we will consider perturbations of the the fluid velocity and temperature. 

In the time-independent case, such perturbations would be Fourier-expanded, and the translation invariance in time $\tau$ and space coordinates $(x_1,x_2,\xi)$ ensures that this simplifies the equations (because, by symmetry, plane waves $e^{-ik\cdot x}$ are solutions of the equations of motion). However, the metric~\eqref{eq:MilneMetric} breaks time translation invariance. Therefore, we apply a \emph{quasi-static approximation}, in which we assume that the time scale of the Bjorken-expansion is much longer than the time scale of perturbations. This restores time-translation invariance on a fixed time slice $\tau=\tau_0$ of the Milne metric~\eqref{eq:MilneMetric}, $g^{\mathrm{(Milne)}}_{\mu\nu}(\tau=\tau_0)$.  
This is similar to a WKB approximation~\cite{wentzel1926verallgemeinerung,kramers1926wellenmechanik,brillouin1926mecanique}, where the potential is assumed to vary slowly along a spatial direction, restoring spatial translation invariance.  

In order to Fourier expand the perturbations we recall that the dot product between two vectors is determined by contraction with the metric, in our case that is the Milne metric~\eqref{eq:MilneMetric} evaluated on a fixed time slice $\tau_0$. This implies that a scalar product of the two four-vectors $x^\mu=(\tau,x_1,x_2,\xi)$ and $k^\nu=(\omega,k_1,k_2,\tilde{k}_\xi)$ is defined by
\begin{equation} 
\label{eq:MilneProduct}
    k\cdot x = k^\mu g^{\mathrm{(Milne)}}_{\mu\nu}(\tau_0) x^\nu = -\omega \tau +k_1 x_1+ k_2 x_2 +\tau_0^2 \tilde{k}_\xi \xi = -\omega \tau +k_1 x_1+ k_2 x_2 +\tau_0 k_\xi \xi \, ,
\end{equation}
where we define the momentum in $\xi$-direction as $k_\xi=\tau_0 \tilde{k}_\xi$ in order to give it the correct dimension. 
The $\tau$ of the first term comes from $x^\mu$, and the $\tau_0$ of the last term stems from $g^{\mathrm{(Milne)}}_{\mu\nu}(\tau_0)$. 
The definition~\eqref{eq:MilneProduct} implies explicit $\tau_0$-dependence for the scalar product appearing in the Fourier transformation. It also offers an interpretation of the quasi-static approximation: assuming that the Fourier modes describing sound perturbations occur on time scales much shorter than the time scale of the Bjorken expansion, we consider the perturbations' momentum vector as dynamical, while the Milne metric describing the Bjorken expansion is frozen in time (on that time slice labeled by $\tau_0$). 

\subsubsection{Transverse modes}
We first consider modes propagating transverse to the Bjorken-expanding direction. 
Mathematically, this corresponds to a first-order perturbation of the fluid four-velocity $u^\mu$,
\begin{equation}
    u^\mu = \left(1,\, \delta u_1(\tau,x_1),\, \delta u_2(\tau,x_1),\, \delta u_\xi (\tau,x_1)\right).
\end{equation}
The corresponding spacelike unit vector $n^\mu$ is
\begin{equation}
    n^\mu = \left(\tau_0 \, \delta u_\xi (\tau,x_1),\, \delta u_2(\tau,x_1),\, -\delta u_1(\tau,x_1),\, 1/\tau_0\right),
\end{equation}
where $\tau_0$ denotes the time slice on which the quasi-static calculation is performed. 
The anisotropic energy–momentum tensor~\eqref{Tconstitutive2} with perturbations takes the form
\begin{equation}\label{Tconstitutive2_Perturbed_trans}
    T^{\mu\nu} =
    (\epsilon+\delta \epsilon)\, u^\mu u^\nu
    + (P_T+\delta P_T)\, \Xi^{\mu\nu}
    + (P_L+\delta P_L)\, n^\mu n^\nu
    + \pi^{\mu\nu}_\perp
    + \pi^{\mu\nu}_\parallel
    + n^{\langle \mu} n^{\nu \rangle} \left( \eta_1 \theta_\perp + \eta_2 \theta_\parallel \right).
\end{equation}

Quasi-statically, the perturbations are assumed to take the plane-wave form
\begin{equation}
    \delta u_i(\tau, x_1) \sim e^{-i \tau \omega + i k x_1}.
\end{equation}
Then the conservation law $\nabla_\mu T^{\mu \nu}=0$, to first order in perturbations, gives\footnote{Notice that we also made approximate use of Eq. (\ref{Isoropic expansion rate}) when deriving Eqs. (\ref{transverse_equation_1}) and (\ref{transverse_equation_2}), since we are studying, in a quasi-static manner, an anisotropic system close to its isotropic equilibrium limit.}
\begin{equation}\label{transverse_equation_1}
    - i k \, \delta u_1 (P_T(\tau_0)+\epsilon(\tau_0)) + i \omega \, \delta \epsilon = 0,
\end{equation}
and
\begin{equation}\label{transverse_equation_2}
    \delta u_1 \left[ 2k^2(2\eta_1(\tau_0)+\eta_2(\tau_0)+6\eta_\perp(\tau_0)) 
    - 9 i \omega (P_T(\tau_0)+\epsilon(\tau_0)) \right] 
    + i k \, \delta P_T = 0.
\end{equation}
Combining Eqs.~(\ref{transverse_equation_1}) and (\ref{transverse_equation_2}) yields
\begin{equation}
    2k^2 \omega \big(2\eta_1(\tau_0)+\eta_2(\tau_0)+6\eta_\perp(\tau_0)\big)
    + 9 i \big(c_\perp(\tau_0)^2 k^2 - \omega^2\big)\big(P_T(\tau_0)+\epsilon(\tau_0)\big) = 0,
\end{equation}
where $c_\perp(\tau_0)^2 \equiv \delta P_T / \delta \epsilon$. In the hydrodynamic limit $\omega, k \to 0$, the dispersion relation is 
\begin{equation}\label{hydro:Transverse_dispersion}
    \omega = \pm c_\perp(\tau_0) k - i \, \frac{6\eta_\perp(\tau_0)+2\eta_1(\tau_0)+\eta_2(\tau_0)}{9\,\left ( P_T(\tau_0)+\epsilon(\tau_0)\right )} \, k^2 + \mathcal{O}(k^3),
\end{equation}
which reproduces the dispersion relation in the isotropic equilibrium limit, since $2\eta_1(\tau_0)+\eta_2(\tau_0) \to 0$ in the isotropic limit as $\tau_0 \to \infty$.

Equation~\eqref{hydro:Transverse_dispersion} gives the form of the dispersion relation of transverse sound modes in Bjorken expanding plasma in the quasi-static approximation. Comparing to the standard form of sound dispersion relations, the imaginary $k^2$-term in Eq.~\eqref{hydro:Transverse_dispersion} is interpreted as the transverse sound attenuation coefficient
\begin{equation}\label{eq:GammaP}
    \Gamma_\perp = \frac{6\eta_\perp+2\eta_1+\eta_2}{9\,\left ( P_T+\epsilon\right )} \, ,
\end{equation}
at any given fixed time $\tau$. Equation~\eqref{eq:GammaP} provides a relation between the transverse sound attenuation $\Gamma_\perp$ on one side and the transverse shear viscosity $\eta_\perp$ as well as viscosities $\eta_1$ and $\eta_2$ on the other side. This relation reduces to the known relation for conformal plasmas, $\Gamma=2\eta/(3(\epsilon +P)$ in the isotropic limit.

\subsubsection{Longitudinal modes}
Next, we study perturbations propagating longitudinal along the Bjorken-expanding direction. In this case,
\begin{equation}
    u^\mu = \left(1,\, \delta u_1(\tau,\xi),\, \delta u_2(\tau,\xi),\, \delta u_\xi (\tau,\xi)\right),
\end{equation}
and
\begin{equation}
    n^\mu = \left(\tau_0 \, \delta u_\xi (\tau,\xi),\, \delta u_2(\tau,\xi),\, -\delta u_1(\tau,\xi),\, 1/\tau_0\right).
\end{equation}
The perturbed energy–momentum tensor takes the same structure as Eq.~(\ref{Tconstitutive2_Perturbed_trans}), with perturbations now depending on $(\tau,\xi)$. The plane-wave ansatz reads
\begin{equation}
    \delta u_i(\tau, \xi) \sim e^{-i \tau \omega + i k \tau_0 \xi}.
\end{equation}
The conservation equations give
\begin{equation}\label{longitudinal_equation_1}
    - i k \tau_0 \, \delta u_\xi (P_L(\tau_0)+\epsilon(\tau_0)) + i \omega \, \delta \epsilon = 0,
\end{equation}
and
\begin{equation}\label{longitudinal_equation_2}
    \tau_0 \delta u_\xi \Big[ 4k^2(2\eta_1(\tau_0)+\eta_2(\tau_0)-3\eta_\parallel(\tau_0)) 
    + 9 i \omega (P_T(\tau_0)+\epsilon(\tau_0)) \Big] 
    - 9 i k \, \delta P_L = 0.
\end{equation}
Combining Eqs.~(\ref{longitudinal_equation_1}) and (\ref{longitudinal_equation_2}) yields
\begin{equation}
    - 4 i k^2 \omega \big(2\eta_1(\tau_0)+\eta_2(\tau_0)-3\eta_\parallel(\tau_0)\big)
    - 9 \big(c_\parallel(\tau_0)^2 k^2 - \omega^2\big)\big(P_L(\tau_0)+\epsilon(\tau_0)\big) = 0,
\end{equation}
where $c_\parallel(\tau_0)^2 \equiv \delta P_L / \delta \epsilon$. In the hydrodynamic limit, $\omega, k \to 0$, we obtain
\begin{equation}\label{hydro:Longitudinal_dispersion}
    \omega = \pm c_\parallel(\tau_0)\, k
    - i\,\frac{2\left[3\eta_\parallel(\tau_0) - 2\eta_1(\tau_0) - \eta_2(\tau_0)\right]}
    {9\left(P_L(\tau_0) + \epsilon(\tau_0)\right)}\, k^2
    + \mathcal{O}(k^3).
\end{equation}
This result again correctly reduces to the corresponding dispersion relation
in the isotropic equilibrium limit. 

Equation~\eqref{hydro:Longitudinal_dispersion} gives the form of the dispersion relation of longitudinal sound modes in Bjorken expanding plasma. Comparing to the standard form of sound dispersion relations, the imaginary $k^2$-term in Eq.~\eqref{hydro:Longitudinal_dispersion} is interpreted as the longitudinal sound attenuation coefficient
\begin{equation}\label{eq:GammaL}
    \Gamma_\parallel = \frac{2}{9}\frac{3\eta_\parallel-2\eta_1-\eta_2}{\,P_L+\epsilon} \, ,
\end{equation}
at any given fixed time $\tau$. Equation~\eqref{eq:GammaL} provides a relation between the transverse sound attenuation $\Gamma_\perp$ on one side and the transverse shear viscosity $\eta_\parallel$ as well as viscosities $\eta_1$ and $\eta_2$ on the other side. This relation reduces to the known relation for conformal plasmas, $\Gamma=2\eta/(3(\epsilon +P)$ in the isotropic limit. 

The effective anisotropic hydrodynamic descriptions of sound propagation given in the dispersion relations transverse Eqs.~(\ref{hydro:Transverse_dispersion}) and longitudinal~(\ref{hydro:Longitudinal_dispersion}) to the Bjorken expansion clearly display anisotropic behavior. The two speeds of sound, $c_\perp$ and $c_\parallel$, are free to take distinct values. The two sound attenuation coefficients $\Gamma_\perp$ and $\Gamma_\parallel$ evidently take distinct values, when we compare their equations~\eqref{eq:GammaP} and \eqref{eq:GammaL}, because $P_L\neq P_T$ in Bjorken expanding plasma and in general $\eta_\perp\neq \eta_\parallel$, and as well $\eta_1\neq 0 \neq \eta_2$.  This will be explicitly seen in the next section, when extracting $c_{\perp,\parallel}$ and $\Gamma_{\perp,\parallel}$ from the holographic results. 

\section{Holographic description}
\label{sec:Holographic description}
In our holographic approach, we perform a numerical computation of the full time-evolution of anisotropic $\mathcal{N}=4$ SYM plasma undergoing Bjorken expansion using the methods of Refs.~\cite{Chesler:2008hg,Chesler:2009cy,Chesler:2010bi,Chesler:2013lia}. On each time-slice of the holographic plasma, we compute the lowest quasinormal  modes, which correspond to the hydrodynamic sound modes~\cite{Kovtun:2005ev}. 
This holographic quasinormal mode data encodes the hydrodynamic dispersion relations~\cite{Kovtun:2005ev}. These dispersion relations are then matched to the anisotropic hydrodynamic description which we phenomenologically constructed, as discussed in Sec.~\ref{sec:hydro}. 
This computation again assumes a \emph{quasi-static approximation}, implying that the time scale on which the plasma changes is much longer than the time scale of perturbations. 
\subsection{Holographic setup}
\label{sec:setup} 
In section~\ref{sec:hydroDispersion}, we studied an effective description—namely, hydrodynamics—for a Bjorken-expanding conformal plasma that is strongly coupled and highly dynamic. As any effective field theory, hydrodynamics alone is not sufficient to determine the values of its parameters, in hydrodynamics those parameters are the transport coefficients. To obtain them, a more fundamental microscopic description of strongly coupled systems is required.
A well known holographic model serves as our microscopic description of choice. 

Holographic duality is a powerful framework that provides a microscopic description of strongly coupled systems by relating $\mathcal{N}=4$ Super-Yang–Mills theory to a classical supergravity theory. On the gravity side, the spacetime dual to a Bjorken-expanding plasma is an inhomogeneous and anisotropic asymptotically Anti–de Sitter~(AdS) geometry~\cite{chesler2010boost,Cartwright:2022hlg}:
\begin{equation}\label{BjorkenMetric}
\mathrm{d}s^2 = 2\,\mathrm{d}r\,\mathrm{d}v - A(v,r)\,\mathrm{d}v^2 + e^{B(v,r)} S(v,r)^2 \left(\mathrm{d}x_1^2 + \mathrm{d}x_2^2\right) + S(v,r)^2 e^{-2B(v,r)} \mathrm{d}\xi^2 .
\end{equation} 
Here, $r$ denotes the bulk radial coordinate of AdS, while $\xi=\tfrac{1}{2}\ln[(t+x_3)/(t-x_3)]$ 
is the rapidity in the longitudinal direction (along the Bjorken expansion). 

In order to compute quasinormal modes, we need to consider metric perturbations around the time-dependent \emph{background metric}~\eqref{BjorkenMetric}. In the time-independent case, such metric perturbations would be Fourier-expanded, and the translation invariance in time $\tau$ and space coordinates $(x_1,x_2,\xi)$ ensures that this simplifies the equations (because, by symmetry, plane waves $e^{-ik\cdot x}$ are solutions of the equations of motion). However, the metric~\eqref{BjorkenMetric} breaks time translation invariance. Therefore, we apply the quasi-static approximation as we did in section~\ref{sec:hydro}. 

In the quasi-static approximation, we divide the geometry into a sequence of time slices, each labeled by $v$. In the asymptotic limit $r \to \infty$, the coordinate $v$ approaches the Milne proper time $\tau$ at the AdS boundary, {i.e.}, $\lim_{r \to \infty} v=\tau$ in the Bjorken case. Thus, the time slices can equivalently be labeled by the boundary proper time $\tau$.\footnote{We set Newton’s constant $G_5=1$ and the AdS radius $L=1$.} We therefore choose to use the proper time $\tau$ to label the time slices in the remainder of this paper. 
To study the transport coefficients holographically, we consider
the Fourier-expanded fluctuations on one fixed time slice $\tau=\tau_0$: 
\begin{equation}
  g_{\mu\nu}(x)=g_{\mu\nu}^{0}(x)+h_{\mu\nu}(x),
\end{equation}
where $g_{\mu\nu}^{0}$ is the \emph{background metric}~(\ref{BjorkenMetric}). 
At each time slice of fixed $\tau$, we neglect the time dependence of the \emph{background metric}\footnote{As long as the time derivatives of the background metric $g^0_{\mu\nu}$ are small compared to the metric, i.e., $|\frac{d g^0_{\mu\nu}}{dv}|\ll |g^0_{\mu\nu}|$, the Einstein equations are approximately satisfied. The better the Einstein equations are satisfied, the more accurate is the quasi-static approximation.} 
and apply the plane wave ansatz on that time slice
\begin{equation}\label{Ansatz}
    h_{\mu\nu}(x) \sim e^{-i k\cdot x} h_{\mu\nu}(k) \, .
\end{equation} 
Neglecting the time-dependence of the background metric implies that when deriving the fluctuation equations in appendix~\ref{Holographic fluctuation equations} from the covariant Einstein equations, we neglect all time derivatives acting on the background metric functions $A$, $B$, and $S$. 
This approximation is valid whenever time-derivatives are small compared to the value of each field. These ratios, $\partial_\tau A/A$, $\partial_\tau B/B$, and $\partial_\tau S/S$ are shown in Fig.~\ref{figDM_M} in appendix~\ref{sec:BjorkenMetric} for times between $\tau T=1.0$ (the time at which most measures indicate local thermal equilibrium~\cite{Cartwright:2022hlg}) and $5.5$. As Fig.~\ref{figDM_M} displays, the largest derivative is $\partial_\tau B/ B\approx 12$\% at the earliest time $\tau T=1.04$, already dropping to less than $3$\% at $\tau T=2.19$. Time-derivatives of $A$ and $S$ are smaller.\footnote{The function $A$ is parametrizing the time-time component of the metric and vanishes around a radial coordinate $u\approx 0.9$, which leads to a divergent behavior for $\partial_\tau A/A$ at that location, while both $\partial_\tau A$ and $A$ are finite. } 
Hence, we conclude that neglecting these time-derivatives is a good approximation at times later than $\tau T \approx 2$ and will lead to errors larger than $12$ \% in the computed transport coefficients at earlier times. This leads to the speeds of sound, attenuations, and relaxation times represented by hollow squares in Fig.~\ref{figBSS}, \ref{figSAT}, and \ref{figRT}. Below, we will discuss how including the time derivatives on each fixed time slice improves the quasi-static approximation, leading to the results represented by the hollow triangles in those same figures.  
Again, similar to the WKB approximation~\cite{wentzel1926verallgemeinerung,kramers1926wellenmechanik,brillouin1926mecanique}, which assumes that the potential varies slowly along the spatial direction, the quasi-static approximation used here assumes that the metric changes slowly along the time direction\footnote{The validity of this assumption is also reflected in the comparison between the metric and its time derivative; see Fig.~\ref{figDM_M} for details.} allowing us to apply the time (translation-invariant) plane wave ansatz (\ref{Ansatz}).

With the quasi-static approximation, the system on a fixed time slice is translation invariant in $v,\, x_1,\, x_2$, and $\xi$.  This allows the Fourier expansion of the metric fluctuations on one fixed time slice of the background metric
\begin{equation}
    h_{\mu\nu}(v,x_1,x_2,\xi,r) \sim e^{-i v \omega +i (k_1 x_1 + k_2 x_2 + \tau_0 k_\xi \xi)} h_{\mu\nu}(\omega,k_1,k_2,k_\xi,r) \, .
\end{equation}
where $\tau_0$ is the proper time, labeling the fixed time slice. 
The bulk coordinate $v$ coincides with $\tau$ in the boundary limit, $\lim\limits_{r\to \infty} v=\tau$. 
We stress again that the background metric breaks time translation invariance, and thus the Fourier expansion performed here introduces a systematic error.

\subsection{Computations}
\label{sec:computations} 
In this subsection, we briefly describe how we solve the metric fluctuation equations to obtain the hydrodynamic dispersion relations for the sound modes. 

We begin with the fluctuation equations given in appendix~\ref{Holographic fluctuation equations}. Those equations describe the metric fluctuations after Fourier expanding them on a fixed time slice at boundary time $\tau=\tau_0$ and with the metric functions $A(v,z),\, B(v,z),\, S(v,z)$ evaluated at that fixed boundary time $\lim\limits_{z\to 0}v=\tau=\tau_0$, where $z=1/r$. 

For each metric fluctuation $h_{\mu\nu}$, ingoing boundary conditions at the (apparent) horizon are required, which in ingoing Eddington-Finkelstein coordinates corresponds to only allowing solution which are regular at the apparent horizon. At the conformal boundary, a vanishing boundary condition is required $h_{\mu\nu}(z=0) = 0$. 
We numerically solve the fluctuation equations with these boundary conditions using a pseudospectral method at several chosen values of the spatial momentum. In the spectral matrix, we encode all dynamical equations. 
The ingoing boundary condition is automatically satisfied through our choice of Chebyshev polynomials as basis, because they are regular everywhere and thus also at the apparent horizon. The boundary condition of vanishing at the AdS boundary is implemented by explicitly replacing the rows in the spectral matrix containing the values of the metric fluctuation at the AdS-boundary with that boundary condition. 

This calculation yields dissipative eigenfunctions with quasinormal frequencies 
as functions of the momentum. Of these quasinormal frequencies, we choose the hydrodynamic ones, namely those for which 
$\lim\limits_{k\to0}\omega(k)=0$~\cite{Kovtun:2005ev}. 
Subsequently, these quasinormal frequencies $\omega(k)$ are fitted to the dispersion relations~\eqref{hydro:Longitudinal_dispersion} and~\eqref{hydro:Transverse_dispersion} in order to extract the speeds of sound and attenuation coefficients. We use Mathematica's function `LinearRegress[...]' for this purpose, and report on the fit quality in appendix~\ref{sec:StandardErrorFit}. 

In the same way, we fit the quasinormal frequency data but now allowing a constant term $C$ in the dispersion relations~\eqref{hydro:Transverse_dispersion} and~\eqref{hydro:Longitudinal_dispersion}. 
Here, $C\neq 0$ is chosen as an indicator for the breakdown of the quasi-static approximation and fit procedure. If the dispersion relation develops a gap, the mode is no longer hydrodynamic, and this occurs around early times $\tau T\approx 1.2$, where the system is far from equilibrium. 

In isotropic equilibrium, the relaxation time appears in the third order of the sound dispersion relation~\cite{Baier:2007ix}, as seen from~\eqref{eq:soundDispersion}. This would require us to push all relations in Sec.~\ref{sec:hydro} to third order, which is beyond the scope of this present work. Instead, here we assume the dispersion at third order to take the form which it satisfies in isotropic equilibrium~\eqref{eq:soundDispersion}, however, we allow for the relaxation time $\tau_\Pi$ to take on different values, namely $\tau_\perp$ when it is associated with the transverse shear relaxation (transverse case) and $\tau_\parallel$ in the longitudinal case; in analogy to $c_{\perp,\parallel}$ and $\Gamma_{\perp,\parallel}$. As  a result, the third order terms are assumed to have the form 
\begin{equation}\label{eq:thirdOrderHydro}
  \omega = \dots \pm\frac{\Gamma_{\perp,\parallel}}{2c_{\perp,\parallel}}\left(c_{\perp,\parallel}^2\tau_{\perp,\parallel}-\frac{\Gamma_{\perp,\parallel}}{2}\right)k^3+\dots \, .
\end{equation}
Relaxation times $\tau_\perp$ and $\tau_\parallel$ are obtained from a fit like above by including these third order terms~\eqref{eq:thirdOrderHydro} into~\eqref{hydro:Transverse_dispersion} and~\eqref{hydro:Longitudinal_dispersion}. 

\paragraph*{Including time-derivative corrections} 
The quasi-static computation can be improved by including time derivatives of the background metric on the time slice at $\tau_0$ on which we perform the Fourier expansion. When deriving the fluctuation equations in appendix~\ref{Holographic fluctuation equations} from the covariant Einstein equations, we previously neglected such time derivatives acting on the background metric functions $A$, $B$, and $S$. 
However, we can choose to retain all such time derivatives $\partial_v A$, $\partial_v B$, $\partial_v S$, $\partial^2_v A$, $\partial^2_v B$, $\partial^2_v S$, yielding a more complicated version of each fluctuation equation in appendix~\ref{Holographic fluctuation equations}, which are available from the authors upon request. 
Physically, including these time-derivatives, provides a correction to the strict quasi-static approximation. Including these time derivatives, the fluctuation equations now retain the information about how the background metric is changing with time at this instant $\tau_0$. The size of the first time derivative of each quantity is shown in Fig.~\ref{figDM_M} at different time slices, showing that these derivatives are sizeable at early times in the Bjorken expansion. 

\paragraph*{Correct asymptotic behavior.}  
As an intermediate check of our results, we compute the quasinormal modes on the known late time metric,
\begin{equation}\label{ASP:metric}
\mathrm{d}s^{2} 
= r^{2}\!\left[
   -\!\left(1-\frac{r_{h}^{4}}{r^{4}}\right)\mathrm{d}v^{2}
   + \mathrm{d}x_{\perp}^{2}
   + v^{2}\mathrm{d}\xi^{2}
  \right]
  + 2\,\mathrm{d}r\,\mathrm{d}v \, ,
\end{equation}
which the full metric~\eqref{BjorkenMetric} asymptotes to~\cite{janik2006asymptotic,Chesler:2009cy}; we refer to this behavior of the metric at asymptotically late time as \emph{asymptotic} behavior. From these quasinormal modes on the asymptotic background metric, we extract the transport coefficients. These asymptotic results are indicated by red circles in Fig.~\ref{figBSS}, \ref{figSAT}, and \ref{figRT}. These figures show that the asymptotic results agree with those obtained from the late-time behavior of the full Bjorken-expanding solution. This confirms that the quasi-static framework captures the correct asymptotic physics of the expanding plasma.

\subsection{Transport coefficients in the transverse direction}
Let us first consider sound modes propagating in the transverse plane, {i.e.,} in the $x_1 x_2$-plane. To be specific, we can set the direction of propagation along $x_1$ direction, then
\begin{equation}
    h_{\mu\nu}(v,x_1,r) \sim e^{-i v \omega +i k_1 x_1} h_{\mu\nu}(\omega,k_1,r) \, .
\end{equation}
With the radial gauge $h_{r\mu}=0$, the first--order fluctuation equations 
decouple into three sectors: the \emph{scalar sector} 
$(h_{x_1 x_1},\, h_{x_2 x_2}+h_{\xi\xi},\, h_{vv},\, h_{v x_1})$, 
the \emph{vector sector} 
$(h_{v x_2},\, h_{x_1 x_2},\, h_{v \xi},\, h_{x_1 \xi})$, 
and the \emph{tensor sector} $(h_{x_2 \xi})$. The fluctuation equations (excluding time derivatives on the background metric) are collected in the appendix~\ref{Holographic fluctuation equations}; fluctuation equations including time derivatives on the background metric can be obtained from the authors upon request.

The transverse speed of sound, denoted by $c_\perp$, is contained in the \emph{scalar sector}. 
Using the spectral method, we numerically compute the quasinormal modes at different momentum values to obtain the corresponding dispersion relation in the hydrodynamic regime with $\omega,k \ll 1$. From this relation, we extract the coefficient of the $\mathcal{O}(k)$ term, which we interpret as the out-of-equilibrium speed of sound $c_\perp$.\footnote{See Appendix~\ref{appedix:holographic results for equilibrium case} for the values of the $\mathcal{N}=4$ SYM speed of sound, sound attenuation coefficient, and the relaxation time for the isotropic equilibrium state.}

As shown in the top panel of Fig.~\ref{figBSS}, the triangles (dots) represent the results obtained with (without) the time-derivative corrections of the metric functions $A(v,r)$, $B(v,r)$ and $S(v,r)$ included in the linear fluctuation equations of $h_{\mu\nu}$. The solid curves correspond to the speed of sound computed from the thermodynamic definition~\cite{Cartwright:2022hlg}.  

The transverse speed of sound, $c_\perp^2$, is observed to asymptotically converge to $1/3$ from above, consistent with the conformal equilibrium value. Physically, this reflects the fact that the transverse geometry is less affected by the rapid longitudinal dilution, so perturbations there retain extra support until the system relaxes. When time-derivative corrections are included (triangles), the results shift closer to the thermodynamic calculation of Ref.~\cite{Cartwright:2022hlg}. The speed of sound data including time-derivatives of the background metric (hollow triangels in Fig.~\ref{figBSS}) remains reliable down to times as early as $\tau T\approx 1.2$, the speed of sound data without those derivatives is unreliable much earlier, highlighting the importance of such corrections in accurately capturing out-of-equilibrium transport. 
We track the breakdown of the quasi-static approximation by allowing an unphysical complex valued constant term in the sound dispersion relations~\eqref{hydro:Transverse_dispersion} and~\eqref{hydro:Longitudinal_dispersion}, and define breakdown of the quasi-static approximation at the time $\tau T$ at which that unphysical constant reaches a magnitude of approximately 10\% of the equilibrium speed of sound, i.e., a value of $0.06$. This is the case at $\tau T\approx 1.4$, as seen from Figs.~\ref{Bjorken_Real_Const_terms} and~\ref{Bjorken_Imaginary_Const_terms}. 

We furthermore applied the quasi-static analysis to the asymptotic spacetime of the Bjorken expansion~\cite{Chesler:2009cy,janikAsymptoticPerfectFluid2006}, described by the metric given in~\eqref{ASP:metric} and found results (red circles in Fig.~\ref{figBSS}) that agree with the late-time behavior of the speed of sound obtained from the full Bjorken-expanding solution. This confirms that the quasi-static framework captures the correct asymptotic physics of the expanding plasma.  

\begin{figure}[htb!]
%
    \begin{subfigure}{0.6\textwidth}
        \includegraphics[width=\textwidth]{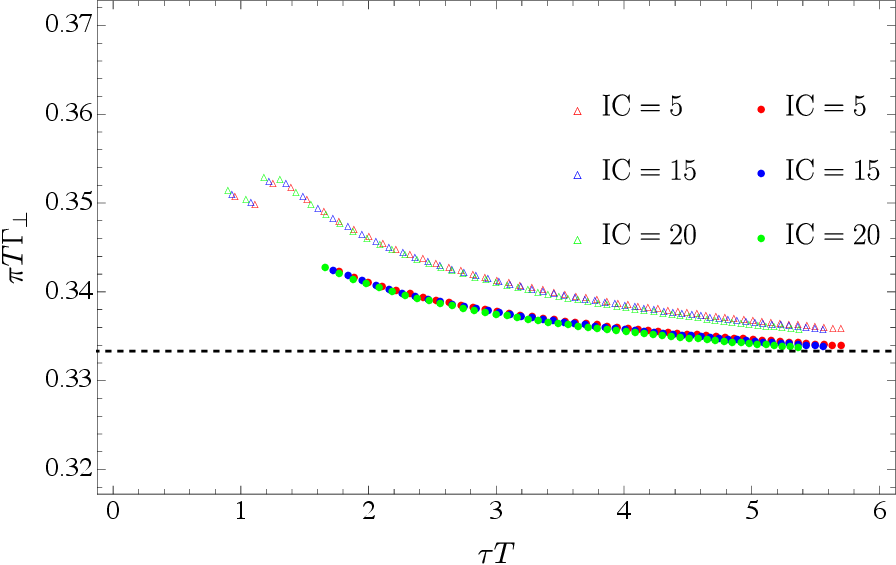}
    \end{subfigure}
     \begin{subfigure}{0.6\textwidth}
        \includegraphics[width=\textwidth]{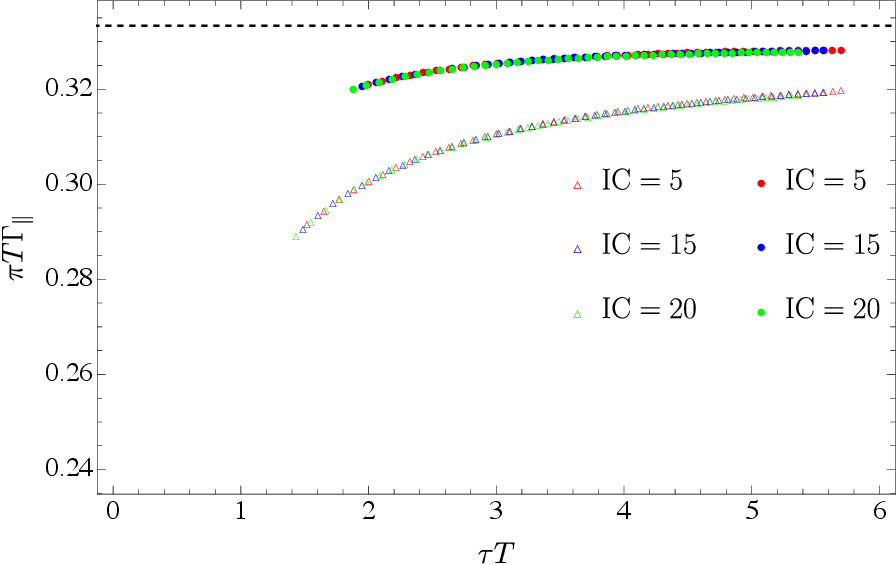}
    \end{subfigure}
    \caption{
    {\it Time evolution of the two sound attenuation coefficients.} The top (bottom) figure shows the transverse (longitudinal) sound attenuation, $\Gamma_{\perp}$ ($\Gamma_{\parallel}$) normalized by a factor $\pi T$. Different colors represent different initial conditions~(IC=5, 15, 20), see table~\ref{tab:IC}. Dashed black lines indicate the isotropic equilibrium value of $\pi T \Gamma=1/3$. The triangles (dots) correspond to calculations with (without) time-derivative corrections. For an estimate of errors, see Figs.~\ref{figSET} and~\ref{figSEL}.  
    }
     \label{figSAT}
\end{figure}

\begin{figure}[htb!]
 \includegraphics[width=0.6\textwidth]{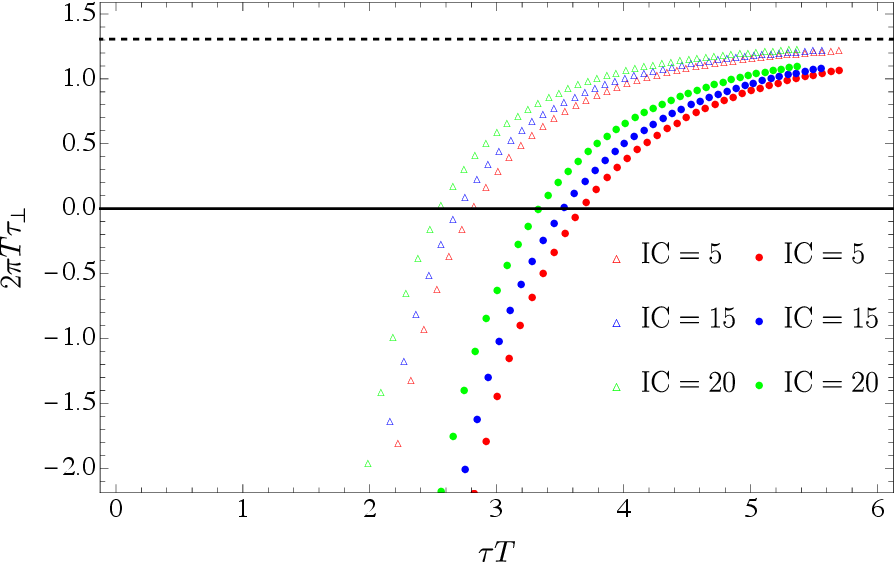}
    \includegraphics[width=0.6\textwidth]{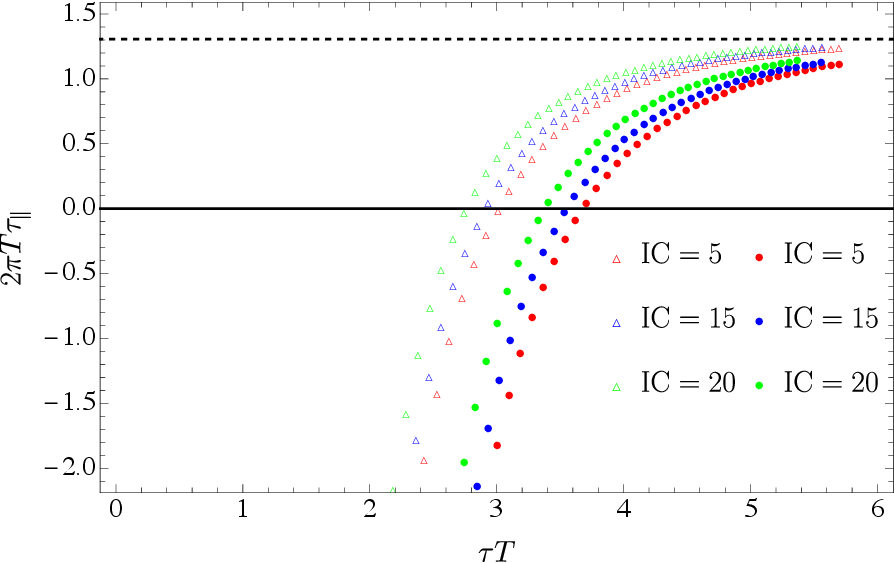}
    \caption{
    {\it Time evolution of the two relaxation time coefficients.} The top (bottom) figure shows the transverse (longitudinal) relaxation time, $\tau{\perp}$ ($\tau{\parallel}$) normalized by a factor $2\pi T$. Different colors represent different initial conditions~(IC=5, 15, 20), see table~\ref{tab:IC}. Dashed black lines indicate the isotropic equilibrium value of $2\pi T \tau=2-\ln 2$. The triangles (dots) correspond to calculations with (without) time-derivative corrections. The quasi-static approximation and fit procedure break down for the relaxation times around $\tau T\lesssim 3$. The quasi-static approximation and fit procedure break down for the sound attenuations including time-derivatives (hollow triangles) around $\tau T\lesssim 1.3$. For an estimate of errors, see Figs.~\ref{figSET} and~\ref{figSEL}. 
    }
     \label{figRT}
\end{figure}

We find the sound attenuation and relaxation time, which appear in the second and third order terms of the dispersion relation. The 
top panel of Fig.~\ref{figSAT} shows that, in the transverse direction, the normalized attenuation $\pi T \Gamma_\perp$ decreases with time and converges to its equilibrium value of $1/3$. This behavior indicates that dissipation for the modes propagating transverse to the Bjorken expansion's direction is initially stronger when the background is rapidly evolving, but weakens as the system expands and gradually loses memory of its non-equilibrium origin. Meanwhile, as shown in the top panel of Fig.~\ref{figRT}, the normalized relaxation time $2\pi T \tau_\perp$ increases from below and approaches its equilibrium value of $2-\ln 2$. This trend suggests that the plasma relaxes perturbations almost instantaneously at early times
\footnote{At even earlier times, the relaxation time becomes negative, signaling the breakdown of the quasi-static method. },
while at later times correlations persist longer and the relaxation slows until the equilibrium value is reached.

\paragraph*{Ranges of validity (transverse direction). }
A few comments on the validity of hydrodynamics, the fit of quasinormal modes to the hydrodynamic dispersion relations, and the quasi-static approximation are in order. 
We mentioned above, that the Bjorken expanding plasma reaches local equilibrium around $\tau T =1$ according to several different measures~\cite{Cartwright:2022hlg}. Therefore, it is reasonable to define local thermodynamic temperature $T(x)$ and fluid velocity $u^\mu(x)$ for times $\tau T > 1$ and we expect hydrodynamics to apply as well, similar to the plasma generated by colliding shocks~\cite{Chesler:2010bi}. 
We expect the speed of sound results to be the most reliable, results for the sound attenuation to be less reliable, while the relaxation times will be the least reliable. This is caused by the fit of the quasinormal mode frequency-momentum data to the hydrodynamic dispersion relation. 
The relaxation time is extracted from a higher order (third order) in the momentum expansion of the frequency, whereas the speed of sound appears at first, and the sound attenuation coefficient at second order. Higher orders in the momentum expansion are more sensitive to numerical errors as well as to systematic errors introduced by the quasi-static approximation. For this reason, the relaxation time depends more strongly on initial conditions than the sound attenuation or speed of sound; this shows in Figs.~\ref{figBSS} and~\ref{figSAT}, where different color data points (different initial conditions) almost coincide, while~\ref{figRT} shows larger differences between data points of different colors.\footnote{Early-time data is known to depend on initial conditions~\cite{Romatschke:2017vte,Heller:2015dha}.} 
Our results are consistent with these expectations, as the speed of sound and the
sound attenuation start showing erratic behavior at $\tau T \approx 1$, see triangles in Figs.~\ref{figBSS} (left) and~\ref{figSAT} (top). Meanwhile, the relaxation times turn negative at $\tau T\approx 2.5$, see triangles in Fig~\ref{figSAT} (top).  So, as expected, the sound attenuation data reaches back to earlier times, $\tau T\approx 1$, while relaxation time data only is starting to become reliable at later times $\tau T> 2.5$. 
Standard errors of the fit are discussed in appendix~\ref{sec:StandardErrorFit} and indicating that it is negligible for $\tau T>1$, with the data becoming unreliable for $\tau T<1$. As a further indicator of validity of the fit and the hydrodynamic dispersion relation, we analyzed the value of putative constant terms in the dispersion relations, which are absent when hydrodynamics applies, as discussed above. Values for these putative constant terms are shown in Figs.~\ref{Bjorken_Real_Const_terms} and~\ref{Bjorken_Imaginary_Const_terms} and should be zero for hydrodynamics to be a good approximation. For the transverse data, the constants are smaller than 10\% of the speed of sound value at times larger than $\tau T= 1$.\footnote{For this estimate, we considered the transverse data for the speed of sound at $\tau T=1$, where $c^2_\perp\approx 0.9$, with the putative constants $C_\perp\approx -0.035+i 0.082$, such that $|C_\perp|/|c_\perp|\approx 0.094$, which is approximately 10\%.} 
This fit rapidly improves as time progresses, and already at $\tau T =2$ the ratio of putative constants to speed of sound value is smaller than 5.5\%.\footnote{For this estimate, we considered the transverse data for the speed of sound at $\tau T=2$, where $c^2_\perp\approx 0.45$, with the putative constants $C_\perp\approx -0.01+i 0.035$, such that $|C_\perp|/|c_\perp|\approx 0.054$, which is approximately 5.4\%.} 

\subsection{Transport coefficients in the longitudinal direction}
Consider the speed of sound $c_\parallel$ along  the direction of the Bjorken expansion, namely $\xi$. Then with Eq.~(\ref{eq:MilneProduct}), the ansatz is
\begin{equation}
    h_{\mu\nu}(v,\xi,r) \sim e^{-i \omega v +i \tau_0 k_\xi \xi} h_{\mu\nu}(\omega,k_1=0,k_2=0,k_\xi,r) \, .
\end{equation}
Here $\tau_0$ is the boundary proper time labeling the time slice on which we perform the quasi-static computation.
Using the ansatz above, the first--order fluctuation equations again 
decouple into three sectors: the \emph{scalar sector} 
$(h_{vv},\, h_{v\xi},\, h_{x_1 x_1}+h_{x_2 x_2},\, h_{\xi\xi})$, 
the \emph{vector sector} 
$(h_{v x_1},\, h_{v x_2},\, h_{x_1 \xi},\, h_{x_2 \xi})$, 
and the \emph{tensor sector} $(h_{x_1 x_2})$.These lengthy fluctuation equations are collected in the appendix~\ref{Holographic fluctuation equations}. 

As before, we use the spectral method to solve numerically the five coupled equations of the scalar sector to find the quasinormal modes and extract the value for $c_\parallel$. The results for $c_\parallel$ at different time slices are shown in the bottom plot of Fig.~\ref{figBSS}. 

In the bottom panel of Fig.~\ref{figBSS}, the squared longitudinal speed of sound, $c_\parallel^2$, initially decreases before turning upward and eventually converging to $1/3$ from below. Compared with the transverse case, the convergence to the equilibrium value occurs more rapidly. This behavior is in line with the expectation that longitudinal modes are more directly influenced by the longitudinal expansion, and thus more quickly driven toward the conformal equilibrium value. Including time-derivative corrections in the quasi-static computation once again shifts the results closer to the thermodynamic calculation of~\cite{Cartwright:2022hlg}. 

The longitudinal sound attenuation, $\pi T \Gamma_\parallel$, shown in the bottom panel of Fig.~\ref{figSAT}, converges to its equilibrium value from below. This indicates that in the out-of-equilibrium regime, longitudinal modes dissipate more slowly than in equilibrium, a feature that can be understood as a stabilizing effect of the strong longitudinal expansion. Meanwhile the normalized relaxation time, $2\pi T \tau_\parallel$, displayed in the bottom panel of Fig.~\ref{figRT}, starts below its equilibrium value and grows toward $2-\ln 2$, following the same qualitative trend as observed for the transverse modes.  

\begin{figure}[htb!]
    \centering
    \begin{subfigure}{0.45\textwidth}
        \includegraphics[width=\textwidth]{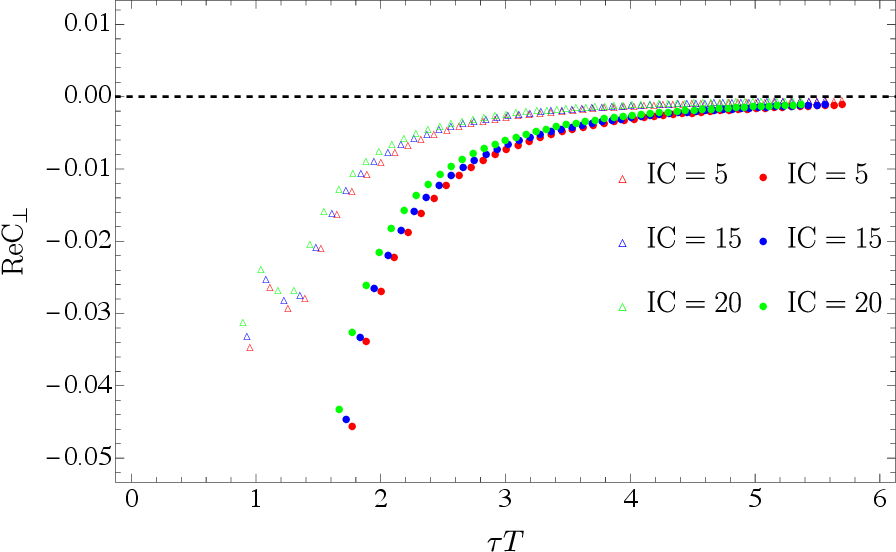}
    \end{subfigure}
    \hfill
     \begin{subfigure}{0.45\textwidth}
        \includegraphics[width=\textwidth]{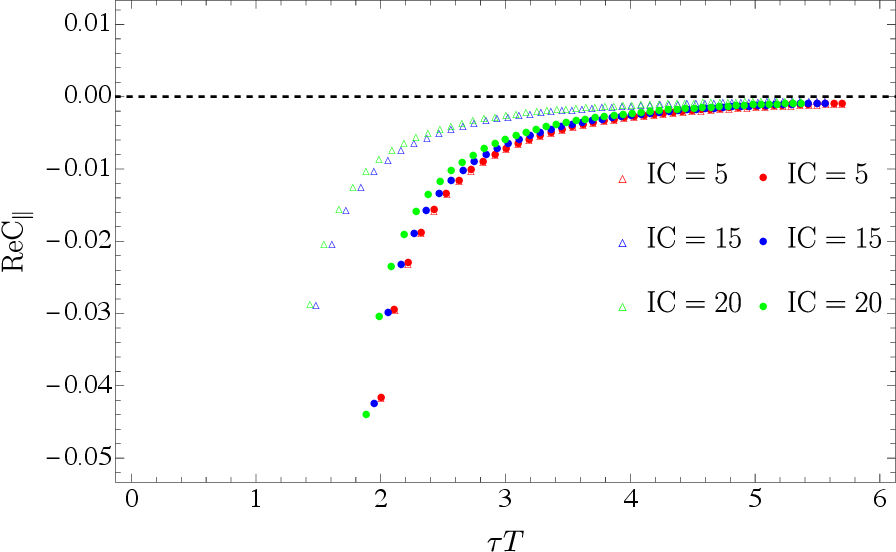}
    \end{subfigure}
    \caption{
    The evolution of the constant terms $\mathrm{Re } C_{\perp,\parallel}$ from the real part of the dispersion relation under different initial conditions(IC). The left (right) plot corresponds to cases in the transverse (longitudinal) direction. Triangle and dot plots are for cases with and without time derivatives, respectively.}
     \label{Bjorken_Real_Const_terms}
\end{figure}

\begin{figure}[htb!]
    \centering
    \begin{subfigure}{0.45\textwidth}
        \includegraphics[width=\textwidth]{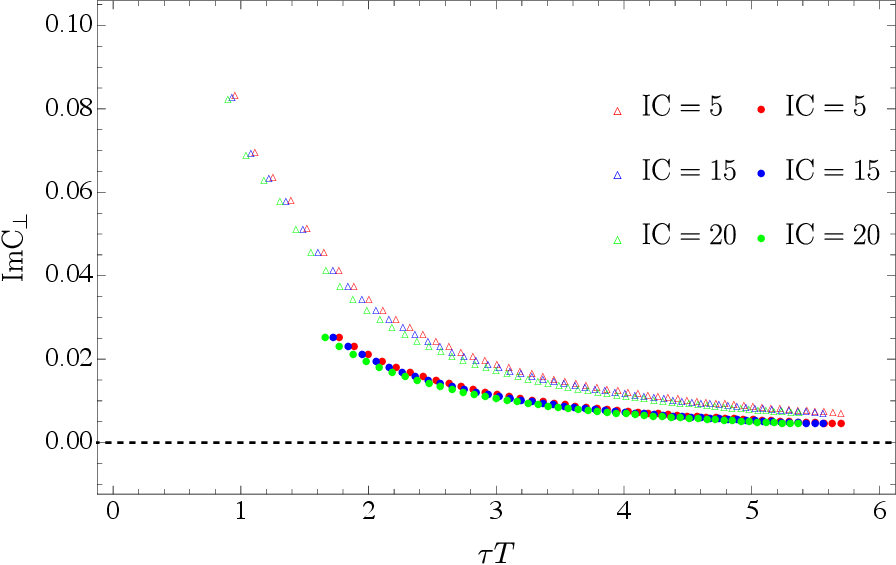}
    \end{subfigure}
    \hfill
     \begin{subfigure}{0.45\textwidth}
        \includegraphics[width=\textwidth]{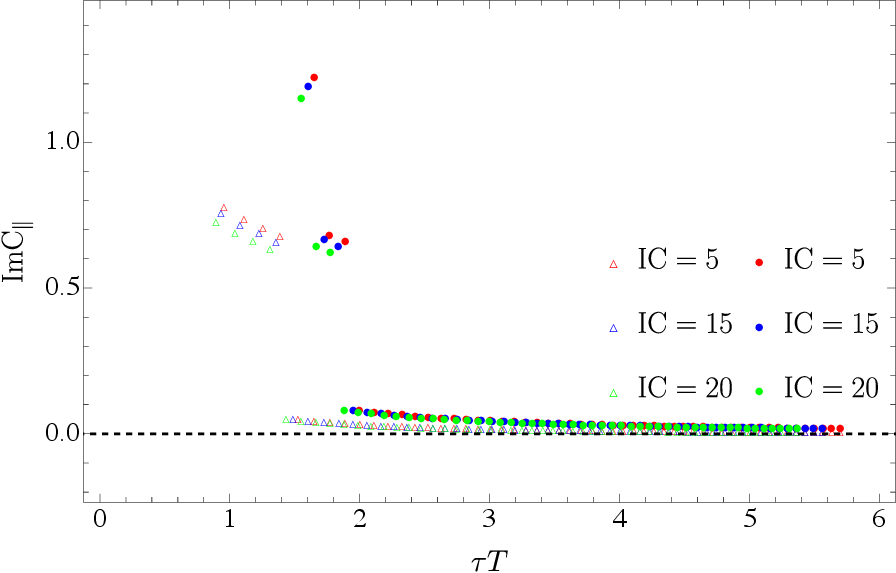}
    \end{subfigure}
    \caption{The evolution of the constant terms $\mathrm{Im} C_{\perp,\parallel}$ from the imaginary part of the dispersion relation under different initial conditions IC. The left (right) plot corresponds to cases in the transverse (longitudinal) direction. Triangle and dot plots are for cases with and without time derivatives, respectively.}
     \label{Bjorken_Imaginary_Const_terms}
\end{figure}

\paragraph*{Ranges of validity (longitudinal direction). }
Similar to the transverse sound modes, also the longitudinal sound modes can only be trusted up to errors stemming from the fit of the hydrodynamic dispersion relation and up to errors from the quasi-static approximation. Standard errors of the fit are discussed in appendix~\ref{sec:StandardErrorFit} and indicating that it is negligible for $\tau T>1$, with the data becoming unreliable for $\tau T<1$. 
The value of putative constant terms in the dispersion relations again serve as indicator for the validity of the hydrodynamic expansion. Values for these putative constant terms are shown in Figs.~\ref{Bjorken_Real_Const_terms} and~\ref{Bjorken_Imaginary_Const_terms} and should be zero for hydrodynamics to be a good approximation. For the longitudinal data, the constants are smaller than 9\% of the speed of sound value at times larger than $\tau T= 1.5$.\footnote{For this estimate, we considered the longitudinal data for the speed of sound at $\tau T=1.5$, where $c^2_\parallel\approx 0.5$, with the putative constants $C_\parallel\approx -0.03+i 0.05$, such that $|C_\parallel|/|c_\parallel|\approx 0.082$, which is approximately 8.2\%.} 
This fit rapidly improves as time progresses, and already at $\tau T =2.5$ the ratio of putative constants to speed of sound value is smaller than 4\%.\footnote{For this estimate, we considered the longitudinal data for the speed of sound at $\tau T=2.5$, where $c^2_\parallel\approx 0.28$, with the putative constants $C_\parallel\approx -0.006+i 0.02$, such that $|C_\parallel|/|c_\parallel|\approx 0.039$, which is approximately 3.9\%.} 

\section{Conclusions}
\label{sec:conclusions}
%
%
In this work, we have computed the propagation of sound modes in a Bjorken expanding plasma by solving the holographically dual metric perturbation equations in a quasi-static approximation, which is valid for perturbations much faster than the Bjorken expansion. We compare our perturbative results to the previous calculation~\cite{Cartwright:2022hlg}, which relied on a thermodynamic definition of the speeds of sound. In our present perturbative definition we do not assume the validity of thermodynamics. This allows us to probe transport properties at times when thermodynamic equilibrium relations are not yet valid.  

Our results show a clear anisotropy between the two sectors. In the late-time regime of the expansion, Bjorken flow suppresses the speed of longitudinal sound modes and reduces their attenuation, while transverse modes exhibit enhanced sound speed and stronger damping relative to equilibrium. Both sectors ultimately converge to the conformal equilibrium values, with longitudinal modes approaching equilibrium more rapidly, in line with the expectation that they are more directly influenced by the expansion dynamics and thus react more immediately pushing back towards equilibrium than the indirectly affected transverse transport.  

Out of equilibrium, a perturbative calculation of sound waves propagating through a time-dependent medium is well-defined and is more reliable than a calculation which relies on the definition $\partial P/\partial \epsilon = c_s^2$, which is strictly true only in global thermodynamic equilibrium. This means that our perturbative results are more reliable than the thermodynamic definition of the speed of sound~\cite{Cartwright:2022hlg}, that is, until the quasi-static approximation breaks down at $\tau T\lesssim 1.4$ (for the transverse speed of sound). The breakdown of the quasi-static approximation is indicated by constant terms in the sound dispersion relation becoming sizable (approximately 10\% of the frequency value), see Figs.~\ref{Bjorken_Real_Const_terms} and~\ref{Bjorken_Imaginary_Const_terms} and discussion in Sec.~\ref{sec:Holographic description}.\footnote{Recall that the bound $\tau T \lesssim 1.4$ applies to the speed of sound, whereas $\tau T \sim 3$
refers to the relaxation time. As discussed in Sec.~\ref{sec:Holographic description}, this discrepancy arises because the speed of sound can be extracted at first order in the momentum expansion, whereas the relaxation time
requires going to third order, and higher-order information is more sensitive to numerical errors, systematic errors, initial data and the quasi-static approximation. } 
The comparison with the thermodynamic definition of the speed of sound, see Fig.~\ref{figBSS}, highlights the role of the time-derivative corrections. Comparing the hollow triangles to the dots in Fig.\ref{figBSS}, incorporating time-derivative corrections systematically brings the perturbative results closer to the thermodynamic results of~\cite{Cartwright:2022hlg}. In addition, the time-derivative corrections allow to extract transport coefficient data starting at earlier times. In particular, for $\tau T \gtrsim 1\,\mathrm{fm}$, when the relation $\epsilon+P = sT$ holds~\cite{Cartwright:2022hlg}, local thermal equilibrium is established\footnote{All measures of local thermodynamic equilibrium analyzed in~\cite{Cartwright:2022hlg} support this statement.}, and the perturbative results including time derivative corrections (hollow triangles) track the hydrodynamic attractor more closely the larger $\tau T$ gets. However, at $\tau T=1.4\,\mathrm{fm}$, the perturbative transverse speed of sound (hollow triangles) deviates by almost a factor of 2 from the thermodynamically defined values (red, blue, green solid curves) and from the sound attractor (black solid curve). \emph{This demonstrates that local thermodynamic equilibrium does not imply that the thermodynamic definition of the speed of sound becomes reliable.}\footnote{This conclusion assumes that our perturbative calculation in this paper, although it contains errors due to the quasi-static approximation, yields a more reliable result for the speed of sound than the thermodynamic definition $c_s^2=(\partial P/\partial \epsilon)$. This assumption is justified because the thermodynamic definition is only strictly valid in global equilibrium, while the perturbative computation of a sound wave traveling through a changing medium is more adequately addressing the non-equilibrium nature of a time-dependent plasma.} 
The agreement between thermodynamically and perturbatively computed longitudinal (transverse) sound speed is good for times later than 2.5 fm/c (3.5 fm/c).\footnote{We adapt the unit matching between gravitational and field theoretic quantities suggested by~\cite{Chesler:2009cy}, where the dimensionless $\tau T=1$ corresponds to $1$ femtosecond divided by the speed of light, 1 fm/c.} %
At earlier times,  $\tau T\lesssim 1.4$, the quasi-static approximation starts breaking down. 
%
Implications for experiments and analysis tools: 
\begin{itemize}
  \item {\bf Measure anisotropic QGP properties: } Our theoretical results provide strong motivation for measuring anisotropic properties of the quark gluon plasma in experiments, because according to our results, transport properties along the anisotropy are significantly different from transport properties transverse to it. 
  \item {\bf Include anisotropies in analysis tools and hydro codes: } Anisotropic plasma properties need to be taken into account in the analysis tools used to evaluate heavy ion collision data. Thus, we strongly advocate to include anisotropic transport coefficients into hydrodynamic simulations used for the analysis of experimental data such as \emph{MUSIC}~\cite{Schenke:2010nt}, \emph{CLVisc}~\cite{Pang:2018zzo}, \emph{VHLLE}~\cite{Karpenko:2013wva}, \emph{Hydro+UrQMD}~\cite{Petersen:2008dd}, as well as in the framework of anisotropic hydrodynamics based on an anisotropic distribution function in kinetic theory referred to as \emph{aHydro}~\cite{Strickland:2014pga,Strickland:2017kux,Alqahtani:2017jwl,Ryblewski:2010bs,Ryblewski:2010tn,Ryblewski:2011aq,Ryblewski:2012rr,Florkowski:2012lba,Bazow:2013ifa}. As the most relevant we suggest including the two shear viscosities, $\eta_\perp$ and $\eta_\parallel$, which appear at first order in the derivative expansion~\eqref{Tconstitutive2}. Whenever sound modes are relevant, Eqs.~\eqref{eq:GammaP} and~\eqref{eq:GammaL} suggest that along with the two shear viscosities $\eta_\perp, \, \eta_\parallel$, also the viscosities $\eta_1,\, \eta_2$ should be taken into account, which all appear at first order in the derivative expansion~\eqref{Tconstitutive2}. At second order in derivatives, $\eta_\perp,\, \eta_\parallel, \, \eta_1, \, \eta_2$ together determine the sound attenuation coefficients $\Gamma_\perp,\, \Gamma_\parallel$ through~\eqref{eq:GammaP} and~\eqref{eq:GammaL}.   
\item {\bf Include anisotropic shear viscosities in future Bayesian analyses: } We suggest taking the two distinct shear viscosities into account in future Bayesian inference analyses such as~\cite{Bernhard:2019bmu,JETSCAPE:2020shq,JETSCAPE:2020mzn}. In addition, the effect of all other anisotropic transport coefficients, such as charge conductivities~\cite{Ghosh:2024fkg,Ghosh:2024owm,Shovkovy:2025yvn}, Hall viscosities~\cite{KKMMT:2026}, as well as sound speed and attenuation studied in this work should be taken into account.
\end{itemize}
%
In light of our theoretical results, we make the following suggestions for the future analysis of experimental data (for example, the data analyzed in~\cite{CMS:2024sgx})
\begin{itemize}
    \item {\bf Focus on non-equilibrium quantities: } One should not try to measure equilibrium observables (like pressure) in a system which is distinctly out of global equilibrium at all times (the Bjorken-expanding plasma generated in heavy-ion collisions displays a pressure anisotropy even at freeze-out and hadronization). Instead, measure dynamical observables, such as transport coefficients (shear viscosities, sound attenuation, and other out-of-equilibrium quantities). 
    \item {\bf Experimental observables for longitudinal vs. transverse directions:} Are there accessible experimental observables, which distinguish between longitudinal and transverse direction compared to the beam line? 
    For example, consider the measurement of longitudinal momentum in the plasma that can possibly be related to longitudinal pressure and other quantities, similar to how the transverse momentum may be related to transverse pressure~\cite{CMS:2024sgx}. 
    Both transverse and longitudinal components of particle momenta are experimentally accessible.
    \item {\bf Time-resolved measurements: }
Performing a truly time-resolved measurement is not feasible, as all information is collected after the ions have collided, hadronized, and frozen out. But would it be possible to perform a time-resolved analysis of the data? For example, photons arriving in the detectors likely stem from early times in the collision, particles with larger cross sections stem from later times. This could reveal the time-dependence of transport quantities, and thus would provide a way of testing our results in this paper. 
    \item {\bf Astrophysical observations and condensed matter experiments: } Other observational systems such as neutron stars with strong magnetic fields (leading to strong anisotropy)~\cite{Kaminski:2014jda}, or out of equilibrium experimental setups should be considered, for example, cold atom clouds or an electron gas, in which anisotropic expansion could be realized~\cite{OHara:2002pqs,Moll_2016}.  
\end{itemize}

As an outlook extending our theoretical work, it is in principle, possible to calculate the dispersion relations of sound modes propagating through the Bjorken expanding background without assuming the quasi-static approximation. This would require a much more involved construction of the complex-valued mode frequencies by a Wigner transformation from a large amount of position space solutions to the position space fluctuation equations as achieved in~\cite{Wondrak:2020tzt,Wondrak:2020kml} for the single decoupled shear fluctuation equation at vanishing momentum. For the sound sector dispersion relations, this procedure is significantly more complex due to the large number of coupled equations to be solved at various nonzero momentum values, and we leave this for future work. 
%
%
\acknowledgments 
{M.K.}~thanks Ulrich Heinz, Igor Shovkovy, Clemens Werthmann for discussions, the Galileo Galilei Institute for Theoretical Physics for the hospitality and the INFN for partial support during the completion of this work. This work was supported, in part, by the U.S.~Department of Energy grant DE-SC0012447. CC was supported in part during the completion of this work by the Netherlands Organisation for Scientific Research (NWO) under the VICI grant VI.C.202.104.

%

\appendix

\section{Properties of the Bjorken expanding metric}
\label{sec:BjorkenMetric}
\begin{figure}
    \centering
    \begin{subfigure}[b]{0.30\textwidth}
    \includegraphics[width=\textwidth]{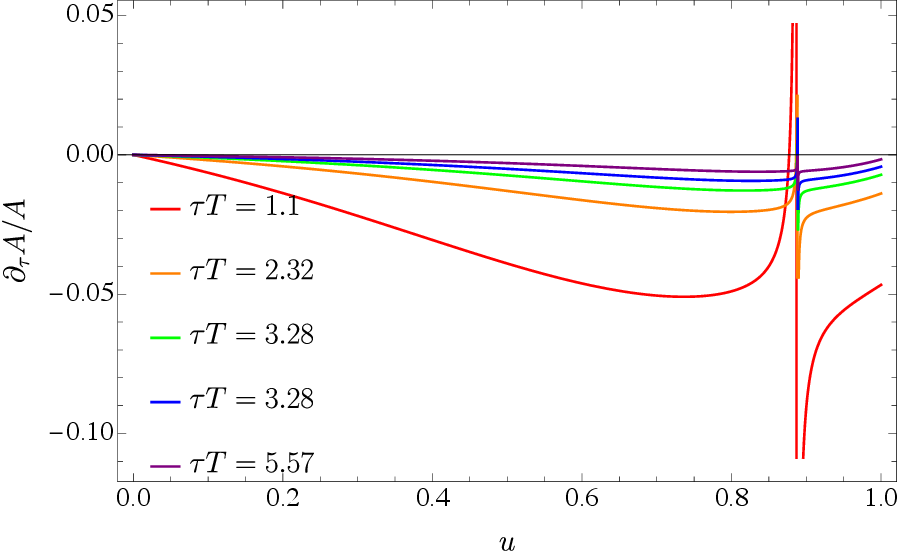}
    \end{subfigure}
    \hfill
    \begin{subfigure}[b]{0.30\textwidth}
    \includegraphics[width=\textwidth]{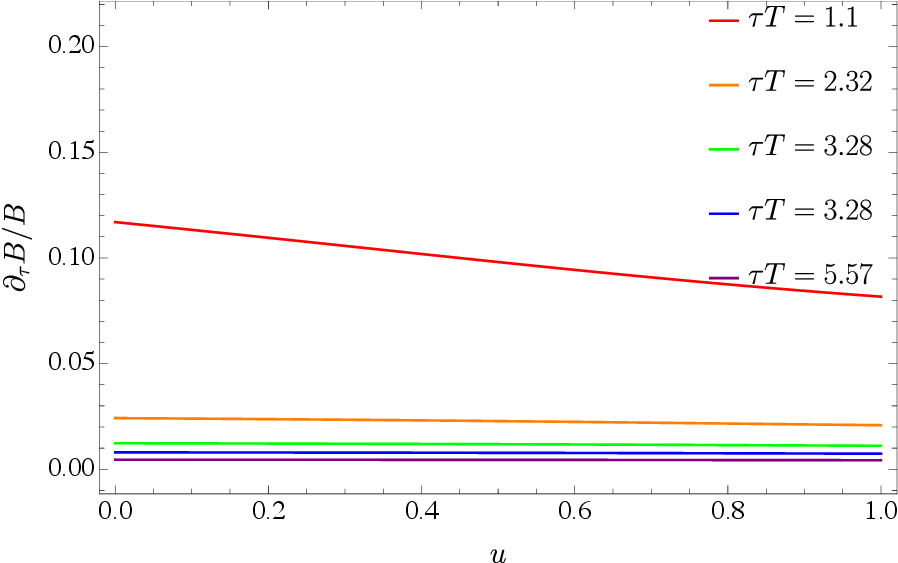}
    \end{subfigure}
    \hfill
    \begin{subfigure}[b]{0.30\textwidth}
    \includegraphics[width=\textwidth]{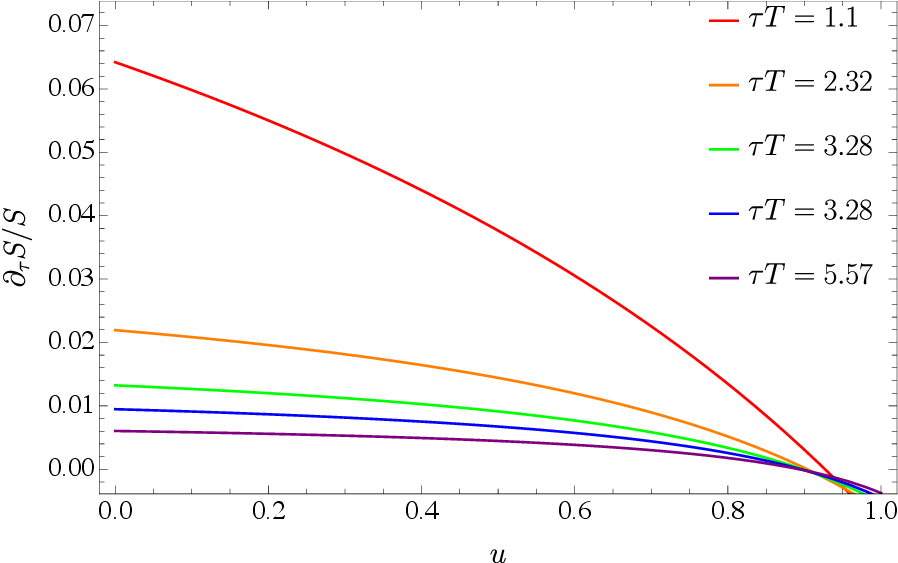}
    \end{subfigure}
    \hfill
    \begin{subfigure}[b]{0.30\textwidth}
    \includegraphics[width=\textwidth]{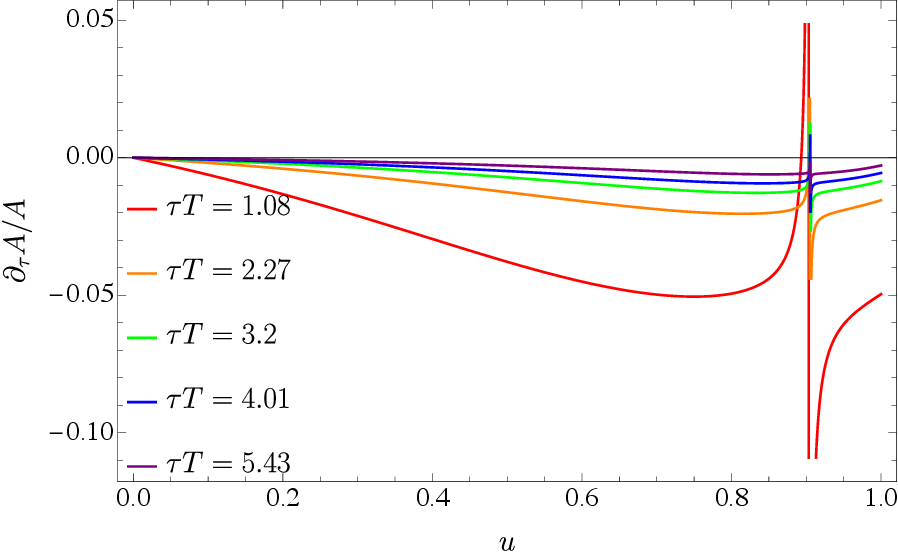}
    \end{subfigure}
    \hfill
    \begin{subfigure}[b]{0.30\textwidth}
    \includegraphics[width=\textwidth]{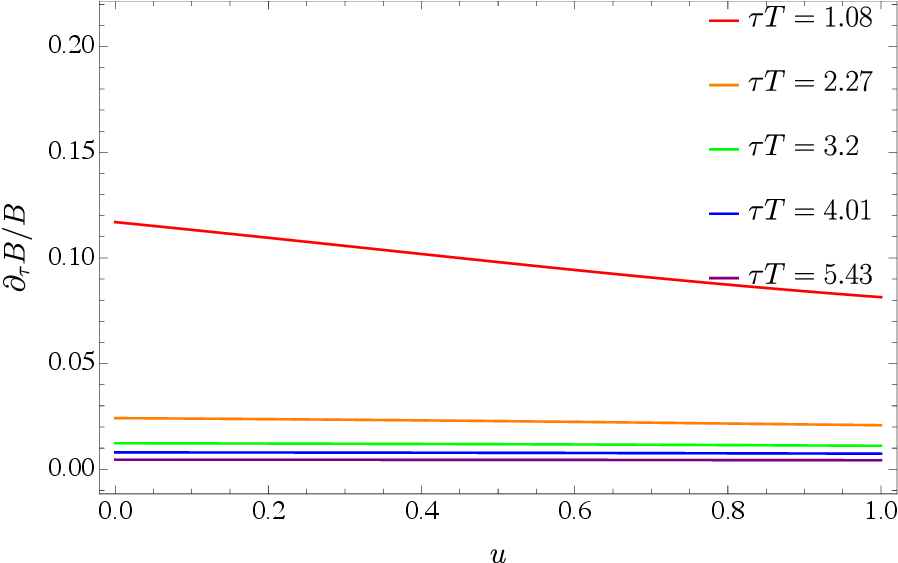}
    \end{subfigure}
    \hfill
    \begin{subfigure}[b]{0.30\textwidth}
    \includegraphics[width=\textwidth]{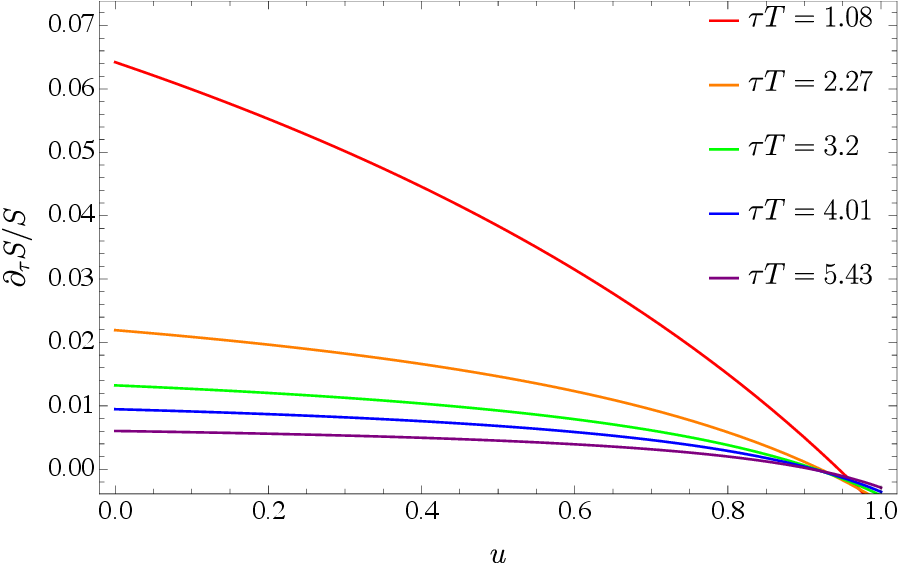}
    \end{subfigure}
    \begin{subfigure}[b]{0.32\textwidth}
    \includegraphics[width=\textwidth]{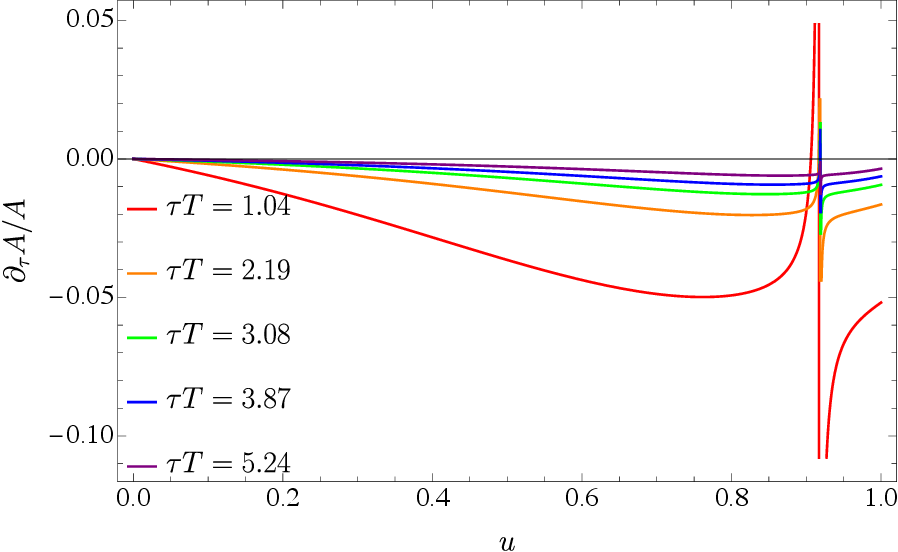}
    \end{subfigure}
    \hfill
    \begin{subfigure}[b]{0.32\textwidth}
    \includegraphics[width=\textwidth]{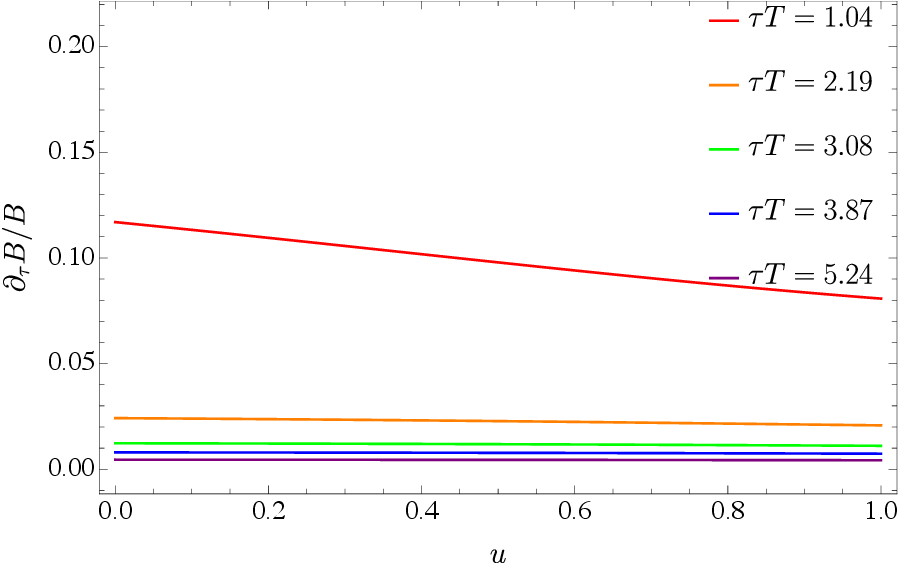}
    \end{subfigure}
    \hfill
    \begin{subfigure}[b]{0.32\textwidth}
    \includegraphics[width=\textwidth]{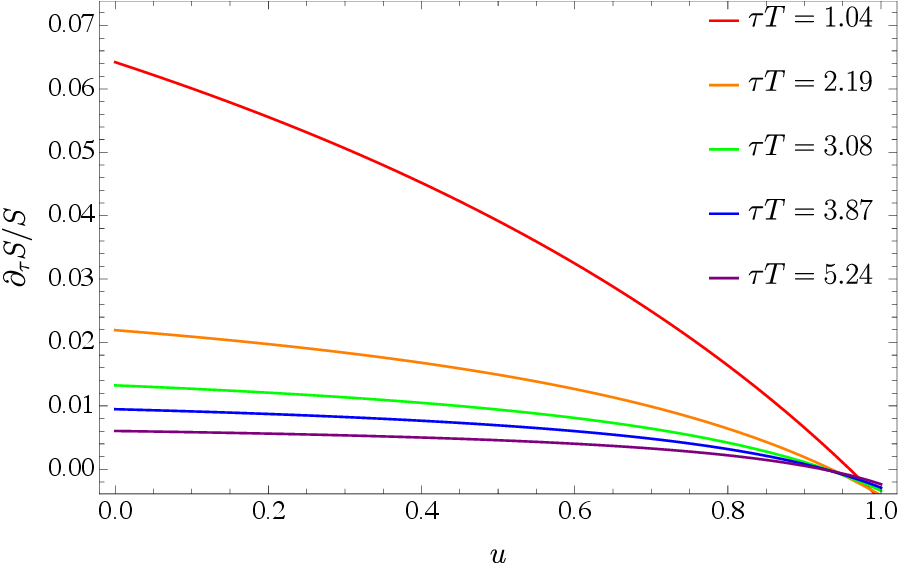}
    \end{subfigure}
    \hfill
    \hfill
    \caption{ \label{figDM_M} 
    Plots of the time derivative of $A$,$B$, and $S$ at different time slices. From the top to the bottom the initial conditions (IC) are 5, 15, and 20, respectively. 
    }
\end{figure}
Fig.\ref{figDM_M} shows the comparison between the metric functions and their time derivatives for the time slices on which we will compute the transport coefficients of the Bjorken expanding plasmas. As illustrated, the time derivatives of the metric functions are significantly smaller than the metric functions themselves in the time region investigated.

\section{Holographic results for an isotropic equilibrium plasma}
\label{appedix:holographic results for equilibrium case}
Compared with the out-of-equilibrium anisotropic case discussed in section~\ref{sec:hydroDispersion}, the hydrodynamic description for an isotropic equilibrium plasma is given by~\cite{Baier:2007ix}
\begin{equation}\label{eq:soundDispersion}
    \omega= C\pm c_{s 0} k - i \frac{\Gamma_{0}}{2} k^2 
    \pm\frac{\Gamma_{0}}{2c_{s 0}}\left(c_{s 0}^2\tau_{\Pi 0}-\frac{\Gamma_{0}}{2}\right)k^3+ O(k^4).
\end{equation}
With $\mathfrak{w}=\omega/(2\pi T)$ and $\mathfrak{q}=k/(2\pi T)$, Eq.~\eqref{eq:soundDispersion} can be rewritten as
\begin{equation}\label{eq:soundDispersion_dimeensionless}
    \mathfrak{w} =\frac{C}{2\pi T}\pm c_{s0} \mathfrak{q} - i\pi T\Gamma_{0} \mathfrak{q}^2 
   \pm\frac{\pi T\Gamma_{0}}{c_{s0}}\left(2\pi T \tau_{\Pi 0} c_{s0}^2-\pi T\Gamma_{0}\right)\mathfrak{q}^3 +O(\mathfrak{q}^4)\, ,
\end{equation}
where the speed of sound, sound attenuation and relaxation time are found to be $c_ {{s0}}^{2}=1/3$ and $\pi T\Gamma_{0} =1/3$, and $2\pi T\tau_{\Pi 0} =2-\text{ln2}$ for the static holographic plasma that is dual to five dimensional Schwarzschild-AdS  black holes with temperature $T$~\cite{Policastro:2002tn,Baier:2007ix}, while a nonzero constant $C$ would indicate that the mode is gapped and therefore non-hydrodynamic by definition, and it is true that $C=0$ in equilibrium. See Figs.~\ref{Bjorken_Real_Const_terms} and \ref{Bjorken_Imaginary_Const_terms} for its out-of-equilibrium behavior.

\section{Initial conditions for solving the bulk metric}
\label{INCs}
In this appendix, we summarize the initial conditions used to solve the bulk metric~(\ref{BjorkenMetric}). 
Since $A$, $B$, and $S$ diverge as $z=1/r\to 0$, we isolate finite parts via the subtraction scheme~\cite{Cartwright:2022hlg}:
\begin{subequations}\label{eq:SubtractionScheme}
 \begin{align}
      A(v,z)&=z^2 A_s(v,z)+\lambda(v)^2+\frac{2\lambda(v)}{z}+\frac{1}{z^2}-2\dot{\lambda}(v),\\
      B(v,z)&=z^4 B_s(v,z)+\Delta_A(v,z),\\
      S(v,z)&=z^3 S_s(v,z)+\frac{v^{1/3}}{z}+\cdots ,
 \end{align}
\end{subequations}
where $\dot{\lambda}\equiv \partial_v\lambda$ and $\lambda(v)$ encodes the residual gauge freedom $r\to r+\lambda(v)$, chosen such that 
\begin{equation}\label{eq:apparent-horizon}
    \dot{S}(v,z_h)=0
\end{equation}
at the apparent horizon.  

To evolve $A_s$, $B_s$, and $S_s$, we require initial data $B_s(v_0,z)$, $\lambda(v_0)$, and the near-boundary coefficient $a_4(v_0)$ (dual to the energy density~\cite{chesler2014numerical}). We parameterize $B_s$ as  
\begin{equation}\label{eq:init_data}
    B_s=\frac{1}{z^4}\Big(B_d+B_\text{AdS}-\alpha\Delta_B\Big), \quad 
    B_\text{AdS}=-\tfrac{2}{3}\log(v+z),
\end{equation}
with
\begin{align}\label{eq:IC_Parameterization}
    B_d&= \sum_{j=1}^{3}\Omega_j z^4 f_j(\gamma_j z)+\sum_{i=0}^5 \beta_i z^{i+4},
\end{align}
where $\alpha$, $\Omega_j$, $\gamma_j$, and $\beta_i$ are free parameters (see Table~\ref{IConditions}).  
\begin{table}[h]\label{IConditions}
 \caption{\textit{Initial Conditions:} The different values of the parameterization of the initial data given in Eq. (\ref{eq:IC_Parameterization}) are displayed. For each parameterization we begin the evolution at $\tau=0.2$ and with initial asymptotic coefficient $a_4=-40/3$ except for initial conditions 24 and 25 for which $a_4=-15.5$ and $a_4=-14.2$ respectively. Note, we do not alter $\alpha$ from the value $\alpha=1$. Doing so leads to initial conditions which do not initially begin as deviations on top of a vacuum AdS solution.  \label{tab:IC}}
 \begin{ruledtabular}
 \begin{tabular}{llllllllllllll}
 IC \# & $\Omega_1$ & $\gamma_1$ & $\Omega_2$ & $\gamma_2$ & $\Omega_3$ & $\gamma_3$ & $\beta_0$ & $\beta_1$ & $\beta_2$ & $\beta_3$ & $\beta_4$ & $\beta_5$ & $\alpha$  \\
1& 0 & 0 & 0 & 0 & 0 & 0 & 0.5 & -0.5 & 0.4 & 0.2 & -0.3 & 0.1 & 1\\
2& 0&0 &0 &0 &0 &0& 0.2 & 0.1 & -0.1 & 0.1 & 0.2 & 0.5 & 1 \\
3&0&0&0&0&0&0& 0.1 & -0.5 & 0.5 &0&0&0& 1\\
4&0&0&0&0&0&0& 0.1 & 0.2 & -0.5 &0&0&0& 1\\
5&0&0&0&0&0&0& -0.1 & -0.4 &0&0&0&0& 1\\
6&0&0&0&0&0&0& -0.2 & -0.5 & 0.3 & 0.1 & -0.2 & 0.4 & 1\\
7&0&0&0&0&0&0& 0.1 & -0.4 & 0.3 &0& -0.1 &0& 1\\
8&0&0&0&0&0&0&0& 0.2 &0& 0.4 &0& 0.1 & 1\\
9&0&0&0&0&0&0& 0.1 & -0.2 & 0.3 &0& -0.4 & 0.2 & 1 \\
10&0&0&0&0&0&0& 0.1 & -0.4 & 0.3 &0& -0.1 &0& 1 \\
11& 1& 1&0&0&0&0&0&0&0&0&0&0& 1\\
12&0&0& 1& 1&0&0&0&0&0&0&0&0& 1\\
13&0&0&0&0&0&0& 0.1 & -0.4 & 0.4 &0& -0.1 &0& 1\\
14&0&0&0&0&0&0& -0.2 & -0.5 & 0.3 & 0.1 & -0.2 & 0.3 & 1 \\
15&0&0&0&0&0&0& -0.2 & -0.3 &0&0&0&0& 1\\
16&0&0&0&0&0&0& -0.2 & -0.5 &0&0&0&0& 1\\
17&0&0&0&0&0&0& -0.1 & -0.3 &0&0&0&0& 1\\
18&0&0&0&0&0&0& -0.1 & -0.2 &0&0&0&0& 1\\
19&0&0&0&0&0&0& -0.5 & 0.2 &0&0&0&0& 1\\
20&0&0&0&0&0&0& -0.2 & -0.4 &0&0&0&0& 1\\
21&0&0&0&0&0&0& -0.2 & -0.6 &0&0&0&0& 1\\
22&0&0&0&0&0&0& -0.3 & -0.5 &0&0&0&0& 1\\
23&0&0&0&0& 1& 8. &0&0&0&0&0&0& 1\\
24& 1& 8 &0&0&0&0& -0.2 & -0.5 &0&0&0&0& 1\\
25&0.5 & 8 &0&0&0&0& -0.2 & -0.5 &0&0&0&0& 1\\
\end{tabular}
\end{ruledtabular}
\label{IConditions}
\end{table}

\section{Fluctuation equations of the scalar sector}
\label{Holographic fluctuation equations}
\subsection{Transverse modes} 
Here we present the scalar (sound) sector of the fluctuation equations for the transverse modes $h_{\mu\nu}(v,x_1,z) \sim e^{-i \omega v +i k_1 x_1} h_{\mu\nu}(\omega,z)$ with $z\equiv1/r$:
\begin{equation}
\begin{aligned}
& 2 k^2 h_{vv} S^2 + 4 k \omega\, h_{vx_1} S^2 + 2 \omega^2 (h_{x_1x_1}+h_{x_2x_2}) S^2 
+ 2 i k z^2 h_{vx_1} S^2 A' + i \omega z^2 (h_{x_1x_1}+h_{x_2x_2}) S^2 A' \\[4pt]
& + z^4 A S^2 (h_{x_1x_1}'+h_{x_2x_2}') A' 
- z^4 A (h_{x_1x_1}+h_{x_2x_2}) S^2 A' B' 
- 2 z^4 A (h_{x_1x_1}+h_{x_2x_2}) S\, A' S' \\[4pt]
& + e^{3B}\!\Big[z^4 A S^2 h_{\xi\xi}' A' 
+ h_{\xi\xi}\!\big(S^2(2 \omega^2 + z^2 A'(i \omega + 2 z^2 A B')) - 2 z^4 A S\, A' S'\big)\Big] \\[4pt]
& - 2 e^{B} S^3\!\Big[z^3 A\big(S(2 h_{vv}' + z h_{vv}'') + 3 z h_{vv}' S'\big) 
+ h_{vv}\big(3 z^2(-i \omega + z^2 A') S' + S(-8 + 2 z^3 A' + z^4 A'')\big)\Big] = 0.
\end{aligned}
\end{equation}
\begin{equation}
\begin{aligned}
& - 2 k \omega\, h_{x_2x_2} S - 2 e^{3B} k \omega\, h_{\xi\xi} S  
+ 2 e^{B} S \Big\{ z^2 S\Big[ S\big(- i k h_{vv}'  
+ i k h_{vv} B' + h_{vx_1}'(- i \omega + z A(-2 + z B'))\big) \\[4pt]
& - z^2 A h_{vx_1}'' + (- i k h_{vv} - z^2 A h_{vx_1}') S' \Big] 
+ h_{vx_1}\Big(S^2(8 + i z^2(\omega + i z^2 A') B') 
+ z^2 S(2 i \omega - 2 z^2 A') S'\Big) \Big\} = 0,
\end{aligned}
\end{equation}
\begin{equation}
\begin{aligned}
& 2 k^2 h_{x_2x_2}  
+ e^{4B} z^2 \Big[ - z^2 A S\, h_{\xi\xi}'(S B' + 2 S') 
+ h_{\xi\xi}\big(S^2 B'(- i \omega - 2 z^2 A B')+\, 2 S(- i \omega - z^2 A B') S' \\[4pt]
& + 4 z^2 A (S')^2 \big) \Big] 
+ 2 e^{3B} k^2 h_{\xi\xi}
- e^{B}\Big\{ h_{x_1x_1}\big(S^2(-16 + z^2 B'(-3 i \omega + z^2 A B')) 
+ 4 z^2 S(z^2 A B' S' + z^2 A (S')^2)\big) \\[4pt]
& + z^2 \Big[ S^2\big(4 i k h_{vx_1}' + 2(z^2 A h_{x_1x_1}'' + h_{x_1x_1}'(2 i \omega + 2 z A + z^2 A')) 
+\, B'(-2 i k h_{vx_1} + z^2 A(-3 h_{x_1x_1}' + h_{x_2x_2}') \\[4pt]
& + i h_{x_2x_2}(\omega + i z^2 A B'))\big) 
+\, 2 S\big(4 i k h_{vx_1} + z^2 A h_{x_2x_2}' + h_{x_2x_2}(i \omega - 2 z^2 A B')\big) S' 
- 4 z^2 A h_{x_2x_2} (S')^2 \Big] \Big\} \\[4pt]
& + 2 e^{2B} z^3 S^2 \Big[ z S\, h_{vv}'(S B' + 2 S') 
+ h_{vv}\big(4 z (S')^2 + S^2(2 B' + z B'') 
+ S((4 + 3 z B') S' + 2 z S'')\big) \Big] = 0,
\end{aligned}
\end{equation}
\begin{equation}
\begin{aligned}
& 2 k^2 h_{x_2x_2} 
+ e^{4B} z^2 \Big[ - z^2 A S\, h_{\xi\xi}'(S B' + 2 S') 
+ h_{\xi\xi}\big(S^2 B'(- i \omega - 2 z^2 A B') + 2 S(- i \omega - z^2 A B') S'  4 z^2 A (S')^2 \big) \Big] \\[4pt]
& + e^{B}\Big\{ h_{x_2x_2}\big(S^2(16 + z^2 B'(3 i \omega - z^2 A B')) 
- 4 z^2 S(z^2 A B' S' + z^2 A (S')^2)\big)+ z^2 \Big[ S^2\big(- i(2 k h_{vx_1}\\[4pt]
&+ \omega h_{x_1x_1}) B' 
+ z^2 A(-2 h_{x_2x_2}'' - h_{x_1x_1}' B' + h_{x_1x_1}(B')^2)+\, h_{x_2x_2}'(-4 i \omega + z(-2 z A' + A(-4 + 3 z B')))\big)  \\[4pt]
& +\, 2 S\big(- 2 i k h_{vx_1} - z^2 A h_{x_1x_1}' + h_{x_1x_1}(- i \omega + 2 z^2 A B')\big) S' 
+ 4 z^2 A h_{x_1x_1} (S')^2 \Big] \Big\}+ 2 e^{2B} z^3 S^2 \Big[ \\[4pt]
& z S\, h_{vv}'(S B' + 2 S') 
+ h_{vv}\big(4 z (S')^2 + S^2(2 B' + z B'') + S((4 + 3 z B') S' + 2 z S'')\big) \Big] = 0,
\end{aligned}
\end{equation}
\begin{equation}
\begin{aligned}
& h_{\xi\xi}\Big(k^2 + e^{B} S^2(8 + z^2 B'(-3 i \omega - 2 z^2 A B')) 
+ 4 z^4 A S B' S' - 2 z^4 A (S')^2 \Big)+ z^2\Big\{ S\big(-2 i k h_{vx_1} \\[4pt]
&  - z^2 A(h_{x_1x_1}'+h_{x_2x_2}') 
+ (h_{x_1x_1}+h_{x_2x_2})(- i \omega - z^2 A B')\big) S' 
+ 2 z^2 A (h_{x_1x_1}+h_{x_2x_2})(S')^2 \Big\} \\[4pt]
& + z^2 S^2\Big[(2 i k h_{vx_1} + i \omega(h_{x_1x_1}+h_{x_2x_2})) B' 
+ z^2 A\big(- e^{3B} h_{\xi\xi}'' + B'(h_{x_1x_1}'+h_{x_2x_2}' 
- (h_{x_1x_1}+h_{x_2x_2}) B')\big) \\[4pt]
&+ e^{3B} h_{\xi\xi}'(-2 i \omega - z(z A' + A(2 + 3 z B'))) \Big] 
+ 4 e^{B} z^2 h_{vv}(S')^2
- 2 e^{B} z S^4\big((2 h_{vv}+ z h_{vv}') B' + z h_{vv} B''\big)\\[4pt]
&- 2 e^{B} z S^3\big((- z h_{vv}'+ h_{vv}(-2 + 3 z B')) S' - z h_{vv} S''\big) = 0,
\end{aligned}
\end{equation}
where $\prime$ denotes $\partial_{z}$.

\subsection{Longitudinal modes}
Here we present the scalar (sound) sector of the fluctuation equations for the longitudinal mode  $h_{\mu\nu}(v,\xi,z) \sim e^{-i \omega v +i \tau_{0} k_\xi \xi} h_{\mu\nu}(\omega,z)$ with $z\equiv1/r$:
\begin{equation}
\begin{aligned}
& A S^2 h_{x_1x_1}' A' z^4 + A S^2 h_{x_2x_2}' A' z^4 - A h_{x_1x_1} S^2 A' B' z^4 - A h_{x_2x_2} S^2 A' B' z^4 - 2 A h_{x_1x_1} S A' S' z^4 \\[4pt]
& - 2 A h_{x_2x_2} S A' S' z^4 + i \omega (h_{x_1x_1}+h_{x_2x_2}) S^2 A' z^2 + 2 \omega^2 (h_{x_1x_1}+h_{x_2x_2}) S^2 + e^{3B}\!\Big(A S^2 h_{\xi\xi}' A' z^4 \\[4pt]
&+ 2 A h_{\xi\xi} S^2 A' B' z^4 - 2 A h_{\xi\xi} S A' S' z^4 + i \omega h_{\xi\xi} S^2 A' z^2 \Big) + e^{3B}\!\Big(2 k^2 h_{vv} S^2 + 2 \omega^2 h_{\xi\xi} S^2+ 2 k h_{v\xi} S^2 (i A' z^2 \\[4pt]
& + 2 \omega)\Big) - 2 e^B S^3 \!\Big(A [S(2 h_{vv}'+ z h_{vv}'')+ 3 z h_{vv}' S'] z^3 + h_{vv}[3(z^2 A'- i \omega) S' z^2 \\[4pt]
&  + S(A'' z^4+ 2 A' z^3 - 8)]\Big) = 0,
\end{aligned}
\end{equation}
\begin{equation}
\begin{aligned}
& 2 e^B z\,[ (2 h_{vv}+ z h_{vv}') B' + z h_{vv} B'' ] S^4 + 2 e^B z\,[ (2 z h_{vv}'+ h_{vv}(3 z B'+4)) S' + 2 z h_{vv} S'' ] S^3 \\[4pt]
& + \Big(8 e^B h_{vv} (S')^2 z^2 - A [2 h_{x_1x_1}''+ B'(h_{x_2x_2}'- h_{x_2x_2} B' + e^{3B}(h_{\xi\xi}'+ 2 h_{\xi\xi} B'))] z^2 \Big) \\[4pt]
& - i (\omega h_{x_2x_2}+ e^{3B}(2 k h_{v\xi}+ \omega h_{\xi\xi})) B' + h_{x_1x_1}'(-4 i \omega - 2 z^2 A' + z A(-4 + 3 z B')) \, S^2 \\[4pt]
& + 2\!\Big(- A h_{x_2x_2}' S' z^2 - 2 e^{3B} i k h_{v\xi} S' + h_{x_2x_2}(2 z^2 A B'- i \omega) S' + e^{3B}[h_{\xi\xi}(-A B' z^2- i \omega)- z^2 A h_{\xi\xi}'] S'\Big) S \\[4pt]
& + 4 z^2 A (h_{x_2x_2}+ e^{3B} h_{\xi\xi}) (S')^2 = 0,
\end{aligned}
\end{equation}

\begin{equation}
\begin{aligned}
& 2 e^B z\,[ (2 h_{vv}+ z h_{vv}') B' + z h_{vv} B'' ] S^4 + 2 e^B z\,[ (2 z h_{vv}'+ h_{vv}(3 z B'+4)) S' + 2 z h_{vv} S'' ] S^3 \\[4pt]
& + \Big(8 e^B h_{vv} (S')^2 z^2 - A [2 h_{x_2x_2}''+ B'(h_{x_1x_1}'- h_{x_1x_1} B' + e^{3B}(h_{\xi\xi}'+ 2 h_{\xi\xi} B'))] z^2 \Big) \\[4pt]
& - i (\omega h_{x_1x_1}+ e^{3B}(2 k h_{v\xi}+ \omega h_{\xi\xi})) B' + h_{x_2x_2}'(-4 i \omega - 2 z^2 A' + z A(-4 + 3 z B')) \, S^2 \\[4pt]
& + 2\!\Big(- A h_{x_1x_1}' S' z^2 - 2 e^{3B} i k h_{v\xi} S' + h_{x_1x_1}(2 z^2 A B'- i \omega) S' + e^{3B}[h_{\xi\xi}(-A B' z^2- i \omega)- z^2 A h_{\xi\xi}'] S'\Big) S \\[4pt]
& + 4 z^2 A (h_{x_1x_1}+ e^{3B} h_{\xi\xi}) (S')^2 = 0,
\end{aligned}
\end{equation}
\begin{equation}
\begin{aligned}
& - 2 k \omega (h_{x_1x_1} + h_{x_2x_2})\, S\, \tau_0 + 2 e^{B} S\Big\{- z^2 S\big[S(z^2 A\, h_{v\xi}'' + i k \tau_0 (h_{vv}'+ 2 h_{vv} B') + h_{v\xi}'(i \omega + 2 z A (1  \\[4pt]
&+ z B')))+ (i k h_{vv} \tau_0 + z^2 A\, h_{v\xi}') S'\big] + 2 h_{v\xi}\big(S^2(4 + z^2(- i \omega + z^2 A') B') + z^2 S (i \omega - z^2 A') S'\big) \Big\} = 0,
\end{aligned}
\end{equation}
\begin{equation}
\begin{aligned}
& e^{2B} k^2 (h_{x_1x_1} + h_{x_2x_2})\, \tau_0^{2} + i \omega z^2 (h_{x_1x_1} + h_{x_2x_2}) S^2 B' + z^4 A S^2 (h_{x_1x_1}' + h_{x_2x_2}') B' - z^4 A (h_{x_1x_1} \\[4pt]
& + h_{x_2x_2}) S^2 (B')^2 - i \omega z^2 (h_{x_1x_1} + h_{x_2x_2}) S S' - z^4 A S (h_{x_1x_1}' + h_{x_2x_2}') S' - z^4 A (h_{x_1x_1} + h_{x_2x_2}) S B' S' \\[4pt]
& +2 z^4 A (h_{x_1x_1}   + h_{x_2x_2}) (S')^2- e^{3B}\Big\{ z^2 S\big[S(z^2 A\, h_{\xi\xi}'' + 2 i k \tau_0( h_{v\xi}' + h_{v\xi} B') + h_{\xi\xi}'(2 i \omega + z(z A' \\[4pt]
&+ A(2 + 3 z B'))))+ 4 i k h_{v\xi} S'\big] + h_{\xi\xi}\big(S^2(-8 + z^2 B'(3 i \omega + 2 z^2 A B')) - 4 z^4 A S B' S' + 2 z^4 A (S')^2\big) \Big\} \\[4pt]
& - 2 e^{B} z^3 S^2 \Big[ z S\, h_{vv}'(S B' - S') + h_{vv}\big(- 2 z (S')^2 + S^2(2 B' + z B'') + S((-2 + 3 z B') S' - z S'')\big) \Big] = 0.
\end{aligned}
\end{equation}

\section{Vaidya AdS spacetime}
\label{sec:SoundVaidya}
In comparison to Bjorken-expanding spacetime, here we present here the study of the asymptotically AdS spacetime \cite{vaidya1951gravitational,javier2010holographic} of the dimension 4+1 for comparison, the metric is
\begin{equation}\label{Vaidya metric}
    \dd s^2 = -\left(r^2 - \frac{m(v)}{r^2}\right) \dd v^2 + 2 \dd r \dd v + r^2 \sum_{i=1}^{3} \dd x_i^2,
\end{equation}
and we choose $m(v)=3-\tanh(v)$, $m(v)=3-\tanh(10v)$, and $m(v)=3-\tanh(100v)$, where $v$ is the Eddington–Finkelstein time, in order to study the spacetime dynamics at different rates. As for Vaidya metric eq.~(\ref{Vaidya metric}), each time slice $v$ can be viewed as a Schwarzschild-AdS metric with temperature $T=m(v)^{1/4}/\pi$ and energy density $\epsilon=m(v)/(4\pi)$~\cite{natsuume2015ads}. 

\begin{figure}
    \centering
    \includegraphics[width=0.5\linewidth]{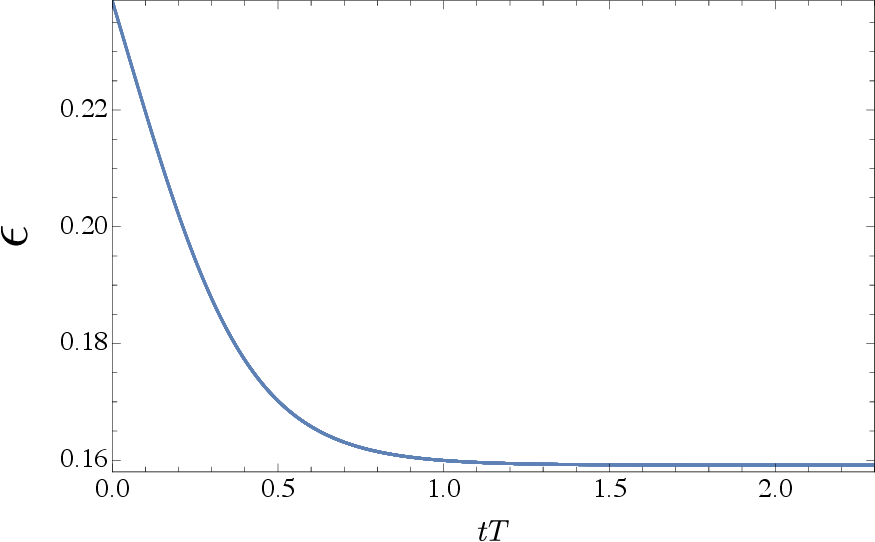}
    \caption{The time evolution of the energy density of Vaidya-AdS spacetime with $m(v) = 3 - \tanh(v)$.}
    \label{Plot_Vaidya_energy_density}
\end{figure}

\begin{figure}
    \centering
    \begin{subfigure}{0.45\textwidth}
        \includegraphics[width=\textwidth]{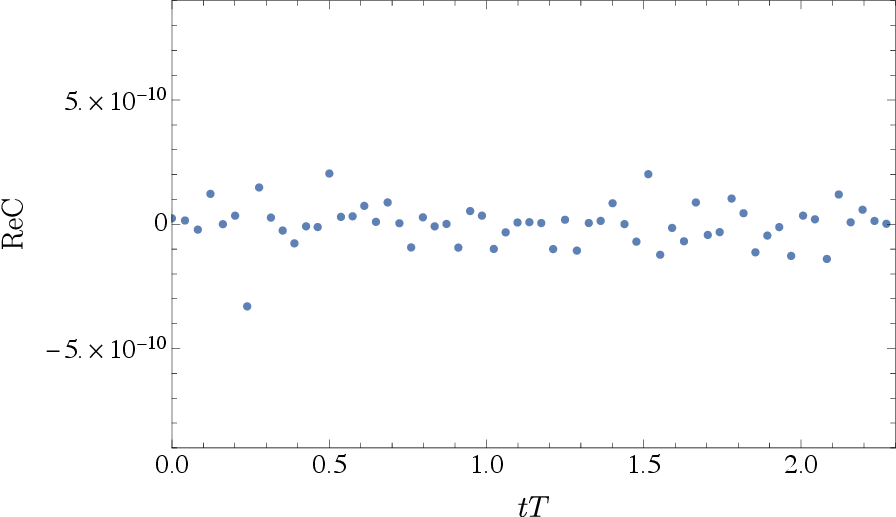}
    \end{subfigure}
    \hfill
     \begin{subfigure}{0.45\textwidth}
        \includegraphics[width=\textwidth]{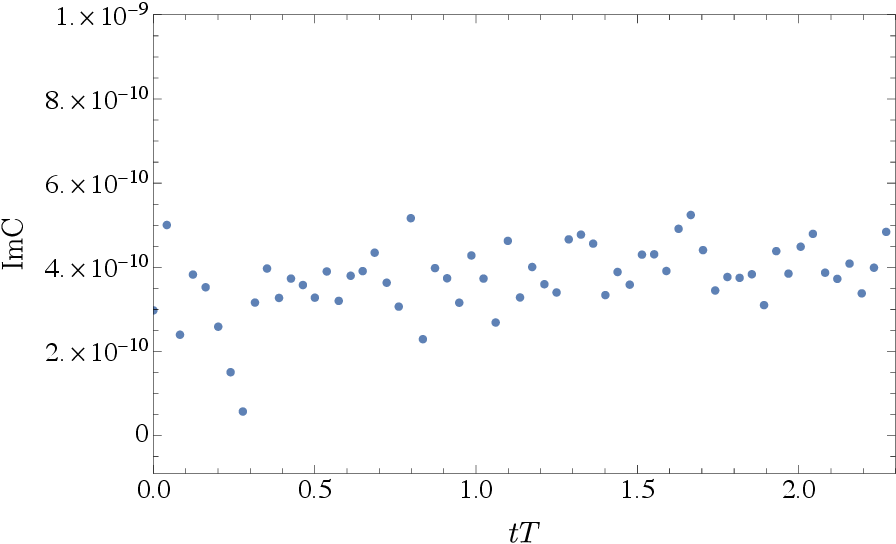}
    \end{subfigure}

    \caption{The time evolution of the constant terms ReC and ImC from the real and imginary parts of the dispersion relation for Vaidya-AdS spacetime.}
     \label{Vaidy_Const_terms}
\end{figure}

\begin{figure}
    \centering
    \includegraphics[width=0.5\linewidth]{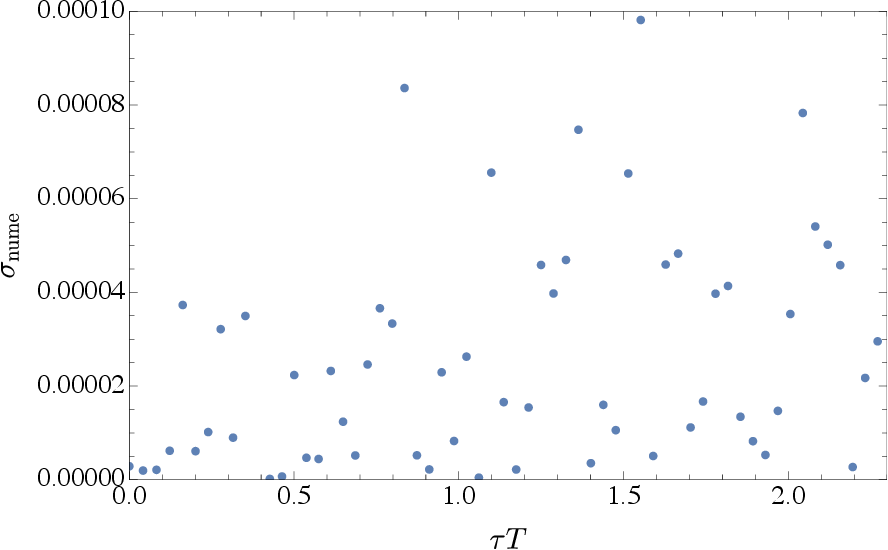}
    \caption{The numerical error at different time slices for $m(v) = 3 - \tanh(v)$.}
    \label{fig:enter-label}
\end{figure}

\begin{figure}
    \begin{center}
    \begin{subfigure}[b]{0.30\textwidth}
    \includegraphics[width=\textwidth]{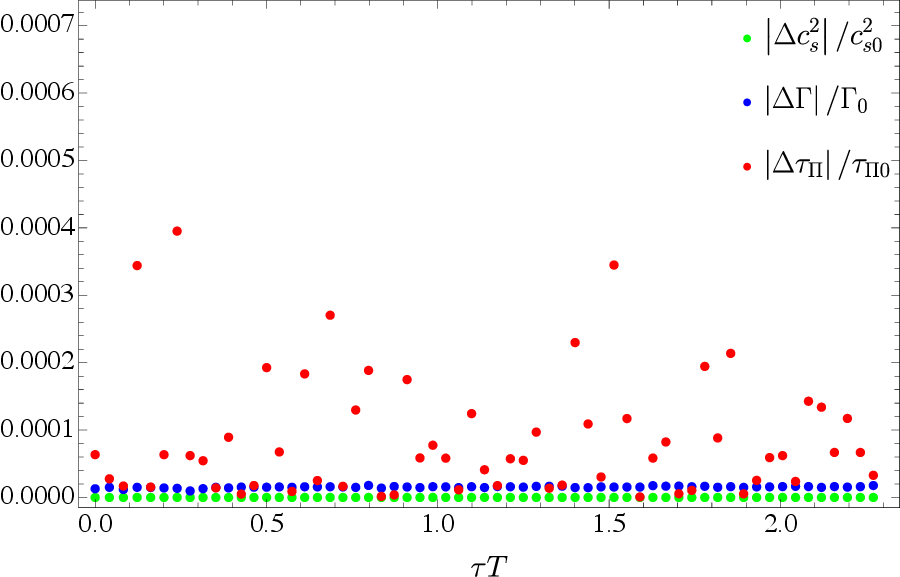}
    \end{subfigure}
    \hfill
    \begin{subfigure}[b]{0.30\textwidth}
    \includegraphics[width=\textwidth]{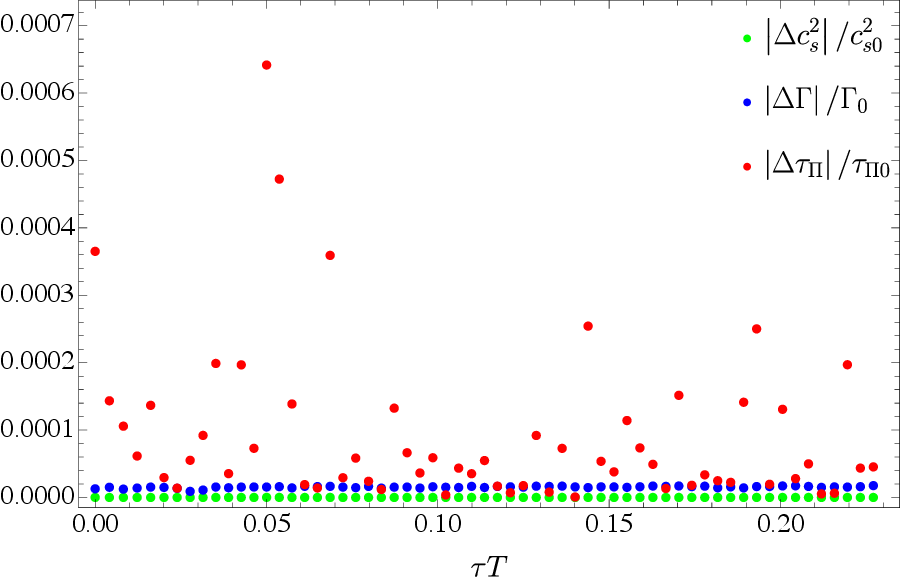}
    \end{subfigure}
    \hfill
    \begin{subfigure}[b]{0.30\textwidth}
    \includegraphics[width=\textwidth]{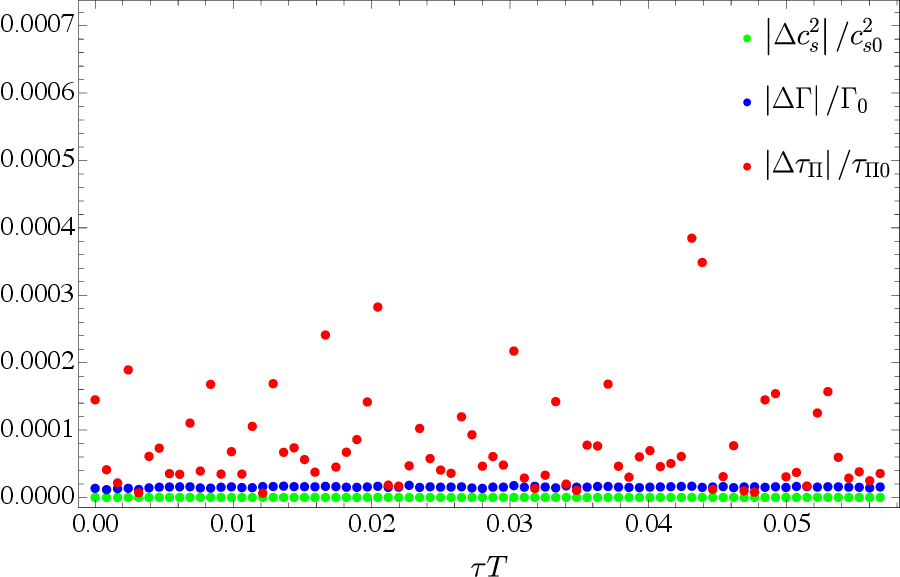}
    \end{subfigure}
    \hfill
    \hfill
      \end{center}
    \caption{The evolution of the speed of sound $c_s$, sound attenuation $\Gamma$, and relaxation time $\tau_{\pi}$ in the Vaidya-AdS spacetime is shown for $m(v) = 3 - \tanh(v)$, $m(v) = 3 - \tanh(10v)$, and $m(v) = 3 - \tanh(100v)$, respectively, from left to right, including their time derivatives. Here, $\Delta c_s^2 = c_s^2 - 1/3$, $\pi T\Delta\Gamma = \pi T\Gamma - 1/3$, and $2\pi T\Delta\tau_{\Pi} = 2\pi T\tau_{\Pi} - (2 - \ln 2)$,  
    while $c^2_{s0}$, $\Gamma_0$, and $\tau_{\Pi 0}$ are the corresponding transport coefficients in equilibrium, see appendix~\ref{appedix:holographic results for equilibrium case}. 
    }
     \label{TCVaidyAdS}
\end{figure}
As shown in Fig.~\ref{TCVaidyAdS}, the transport coefficients (speed of sound, sound attenuation, and relaxation time) in the Vaidya–AdS spacetime, obtained from quasi-static computations, remain constant in time. This indicates that the dynamics of the Bjorken-expanding spacetime are more nontrivial than those of the Vaidya–AdS case.

\section{Standard error of the fit}
\label{sec:StandardErrorFit}
In this section, we present the standard error $\sigma$ of our fitting. Specifically, based on the dispersion relation Eq.(\ref{eq:soundDispersion_dimeensionless}), we perform polynomial fitting using $a_1+a_2x+a_3x^3$ for the real part and $a_4+a_5x^2$ for the imaginary part of the quasi-normal modes at different values of momentum. 
The Standard Error of the Fit $\sigma$ we use defined as
\begin{equation*}
    \sigma = \sqrt{\frac{\sum (y_i - \hat{y}_i)^2}{n - p}},
\end{equation*}
where $y_i$ and $\hat{y}_i$ are values of the data points and fitting functions, respectively. $n$ is the number of data points, and $p$ is the number of parameters of the fitting function.

\begin{figure}
     \centering
     \begin{subfigure}[b]{0.45\textwidth}
        \includegraphics[width=\textwidth]{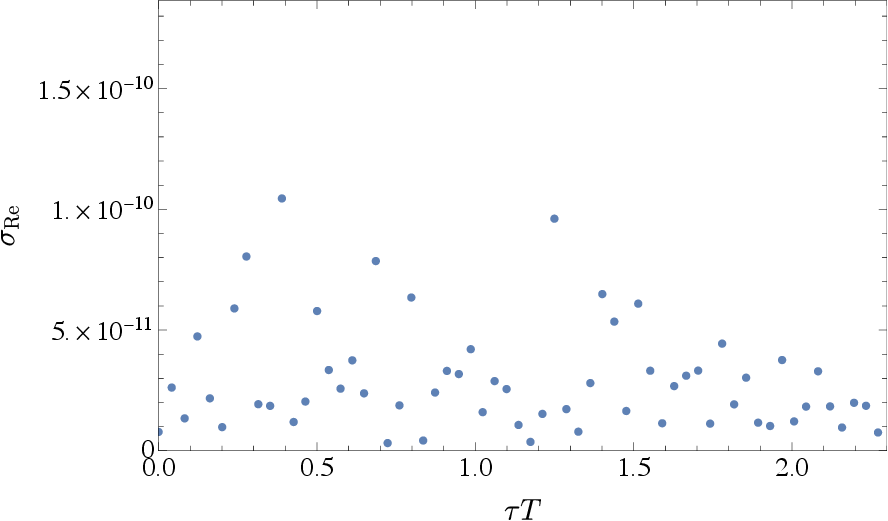}
     \end{subfigure}
     \hfill
     \begin{subfigure}[b]{0.45\textwidth}
         \includegraphics[width=\textwidth]{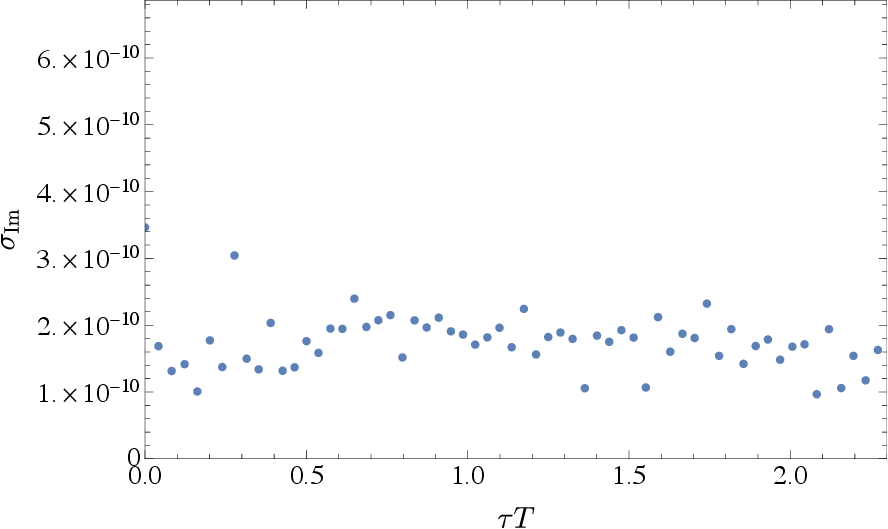}
     \end{subfigure}
     \hfill
         \caption{Standard errors from the the fit for the transverse transport coefficients in the Vaidya case at different time slices. Left: The standard error obtained from fitting the real part of the quasi-normal modes. Right: The standard error obtained from fitting the imaginary part of the quasi-normal modes.}
    \label{figSEV}
\end{figure}

\begin{figure}
     \centering
     \begin{subfigure}[b]{0.45\textwidth}
        \includegraphics[width=\textwidth]{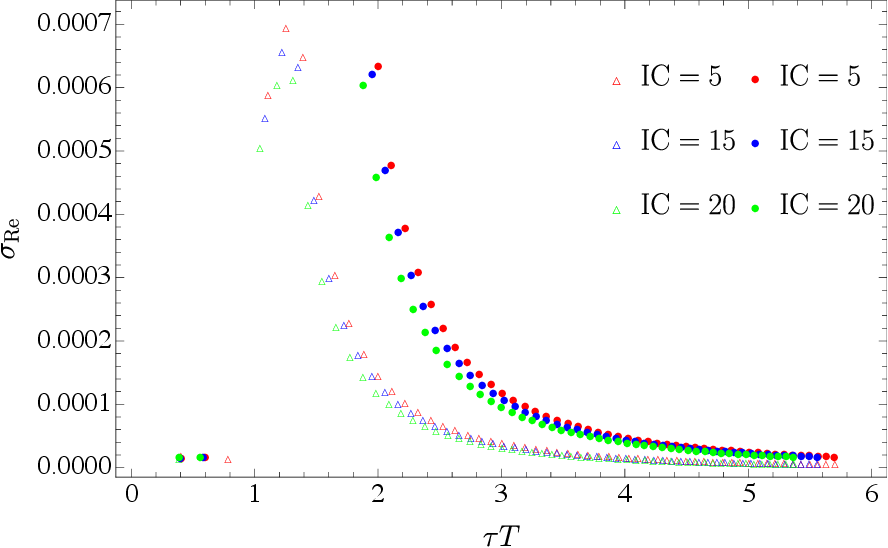}
     \end{subfigure}
     \hfill
     \begin{subfigure}[b]{0.45\textwidth}
         \includegraphics[width=\textwidth]{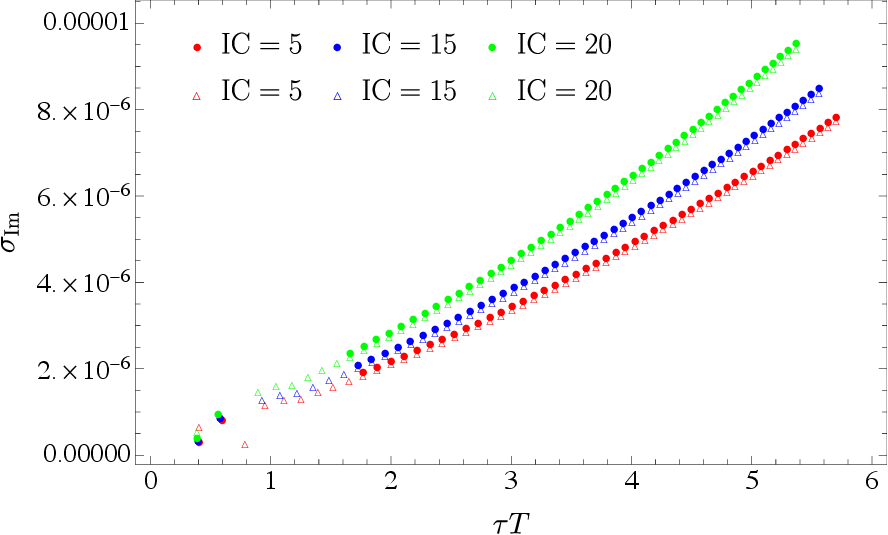}
     \end{subfigure}
     \hfill
         \caption{Standard errors from the the fit for the transverse transport coefficients in the Bjorken expansion at different time slices. Left: The standard error obtained from fitting the real part of the quasi-normal modes. Right: The standard error obtained from fitting the imaginary part of the quasi-normal modes.}
    \label{figSET}
\end{figure}

\begin{figure}
     \centering
     \begin{subfigure}[b]{0.45\textwidth}
        \includegraphics[width=\textwidth]{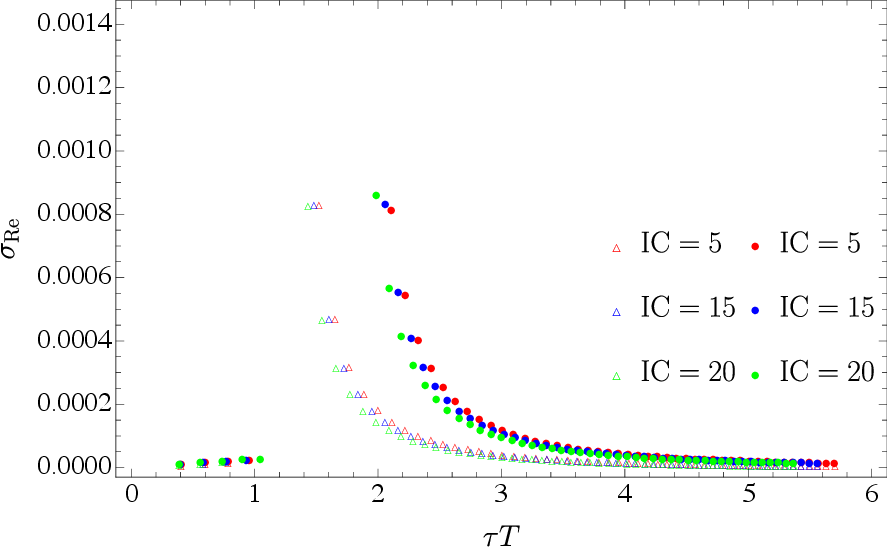}
     \end{subfigure}
     \hfill
     \begin{subfigure}[b]{0.45\textwidth}
         \includegraphics[width=\textwidth]{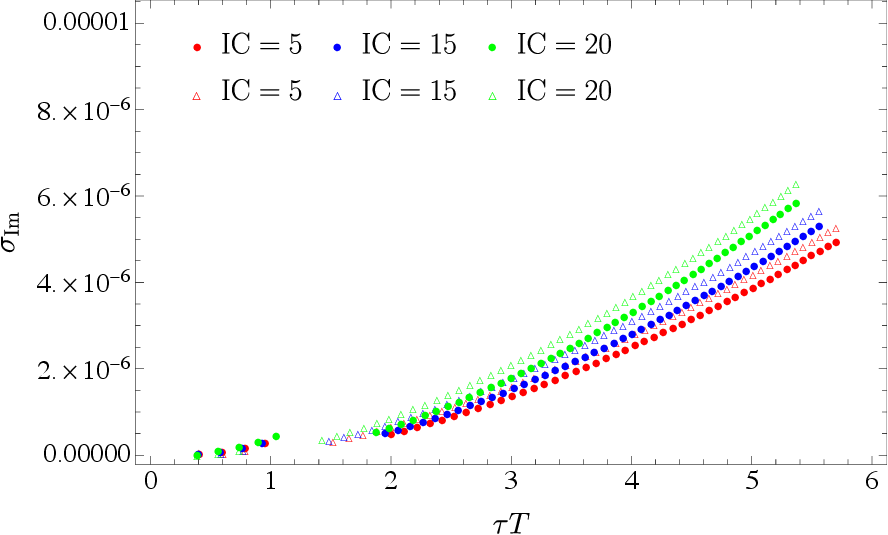}
     \end{subfigure}
     \hfill
         \caption{Standard errors from the fit for the longitudinal transport coefficients in the Bjorken expansion at different time slices. Left: The standard error obtained from fitting the real part of the quasi-normal modes. Right: The standard error obtained from fitting the imaginary part of the quasi-normal modes.}
    \label{figSEL}
\end{figure}

The standard errors of the fitting at different time slices, for Vaidya case, are presented in FIG.\ref{figSEV}, while those for Bjorken expansion are shown in FIG.\ref{figSET} and FIG.\ref{figSEL} for the transverse and longitudinal transport coefficients, respectively.

\end{document}